\documentclass{vldb}
\usepackage{graphicx}
\usepackage{balance}  
\usepackage{epstopdf}
\usepackage{times}
\usepackage{url}
\usepackage{amssymb}
\usepackage{amsmath}
\usepackage{pifont}
\usepackage{cases}
\usepackage{subfigure}
\usepackage{algorithm}
\usepackage{algorithmicx}
\usepackage{bbm}
\usepackage[noend]{algcompatible}
\usepackage{multirow}
\usepackage{enumitem}
\usepackage{rotating}

\newcommand{\boldphi}{\boldsymbol{\phi}}
\newcommand{\boldtheta}{\boldsymbol{\theta}}
\newcommand{\boldpi}{\boldsymbol{\pi}}
\newcommand{\boldeta}{\boldsymbol{\eta}}
\newcommand{\boldnu}{\boldsymbol{\nu}}
\newcommand{\boldlambda}{\boldsymbol{\lambda}}
\newcommand{\bolddelta}{\boldsymbol{\delta}}

\newtheorem{definition}{Definition}

\newtheorem{problem}{Problem}

\newcommand{\eat}[1]{}

\begin{document}


\title{From Community Detection to Community Profiling}



%
%
%
%

\numberofauthors{1} 
\author{%
{Hongyun Cai{\small $~^{{\dagger}{\ddagger}}$}, Vincent W. Zheng{\small $~^{{\dagger}}$}, Fanwei Zhu{\small $~^{\#}$}, Kevin Chen-Chuan Chang{\small $~^{\diamond}$}, Zi Huang{\small $~^{{\ddagger}}$} }%
\vspace{1.6mm}\\
\affaddr{$^{\dagger}$\, Advanced Digital Sciences Center, Singapore}~~~
\affaddr{$~^{{\ddagger}}$\, School of ITEE, The University of Queensland, Australia}  \vspace{1mm} \\
\affaddr{$~^{\#}$\, Zhejiang University City College, China} \hspace{1.8cm}
\affaddr{$^{\diamond}$\, University of Illinois at Urbana-Champaign, USA}  \vspace{1mm} \\
\affaddr{\{hongyun.c, vincent.zheng\}@adsc.com.sg, zhufanwei@zju.edu.cn, kcchang@illinois.edu, huang@itee.uq.edu.au} \vspace{5mm} \\
}

\eat{\author{
%
%
\alignauthor
Hongyun Cai\titlenote{This work is done in Advanced Digital Sciences Center.}\\
       \affaddr{School of ITEE, The University of Queensland, Australia}\\
       \email{h.cai2@uq.edu.au}
\alignauthor
Vincent W. Zheng\\
       \affaddr{Advanced Digital Sciences Center, Singapore}\\
       \email{vincent.zheng@adsc.com.sg}
\and
\alignauthor 
Fanwei Zhu\\
       \affaddr{Zhejiang University City College, China}\\
       \email{zhufanwei@zju.edu.cn}
\alignauthor 
Kevin Chen-Chuan Chang\\
       \affaddr{University of Illinois at Urbana-Champaign, USA}\\
       \email{kcchang@illinois.edu}
\alignauthor 
Zi Huang\\
       \affaddr{School of ITEE, The University of Queensland, Australia}\\
       \email{huang@itee.uq.edu.au}
}
}

\date{1 August 2016}

\maketitle

\begin{abstract}
Most existing community-related studies focus on detection, which aim to find the community membership for each user from user friendship links. However, membership alone, without a complete profile of what a community is and how it interacts with other communities, has limited applications. This motivates us to consider systematically profiling the communities and thereby developing useful community-level applications. In this paper, we for the first time formalize the concept of community profiling. With rich user information on the network, such as user published content and user diffusion links, we characterize a community in terms of both its internal content profile and external diffusion profile. The difficulty of community profiling is often underestimated. We novelly identify three unique challenges and propose a joint Community Profiling and Detection (CPD) model to address them accordingly. We also contribute a scalable inference algorithm, which scales linearly with the data size and it is easily parallelizable. We evaluate CPD on large-scale real-world data sets, and show that it is significantly better than the state-of-the-art baselines in various tasks. 
\end{abstract}


\section{Introduction} \label{sec:introduction}

Thanks to the pioneer studies on community detection \cite{LeskovecLM10,WangWYZ15}, we have been able to model a community in terms of its member users. Such community membership assists us to better understand the network structure. 
However, membership alone, without knowing \emph{what a community is} and \emph{how it interacts with others}, has only limited applications-- e.g., we cannot rank communities by desired characteristics, exploit inter-community diffusions, and visualize communities and their interactions.
With this critical lacking of community ``understanding'', this paper proposes systematic \emph{community profiling}-- to characterize the intrinsic nature and extrinsic behavior of a community-- thereby enabling useful community-level applications. 
As social networks increasingly capture more and richer user information, it is now feasible to profile communities.
E.g., beyond traditional \emph{friendship links} which connect users on a social network, there are also users' attributes, published content, diffused content and so on. 
We can leverage such rich user data to estimate the community profiles. 


\begin{figure*}[t]
  \centering
    \includegraphics[width=1\textwidth]{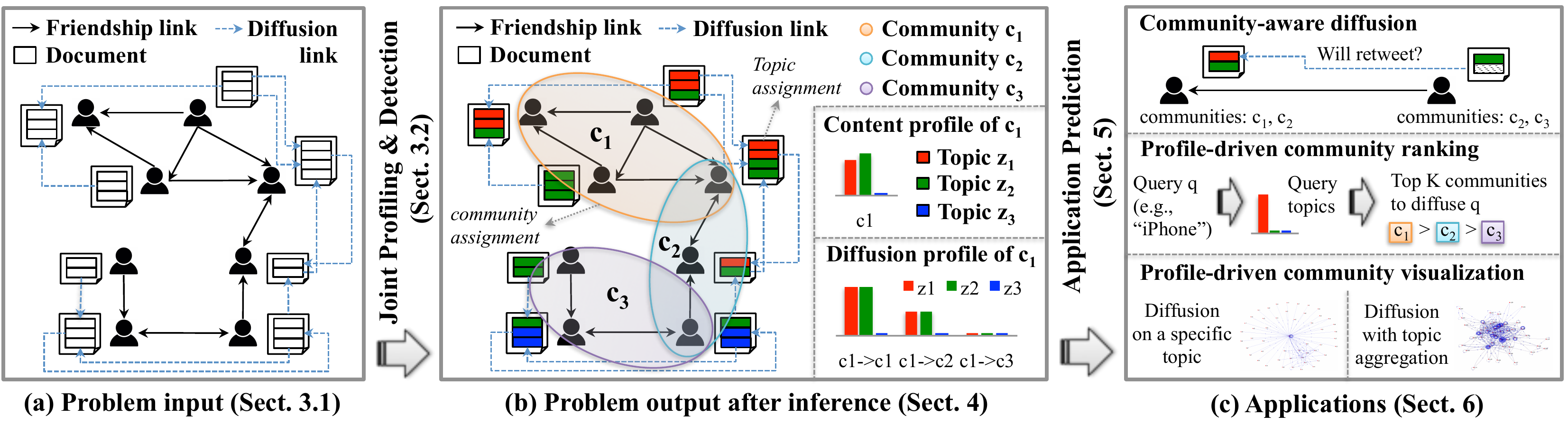}
    \vspace{-3mm}
  \caption{The framework of joint community profiling and detection.}
  \label{fig:framework}
\end{figure*}


In this paper, we for the first time formalize the concept of ``community profile''. We ask two fundamental questions:

\vspace{-2mm}
{\flushleft $\bullet$ {\it What is a community profile}?} 
 By name, the profile should characterize a community, both internally (i.e., what it is) and externally (i.e., how it interacts with others). 
Since a community is an aggregation of users, its profile is essentially an aggregation of user information. 
Denote $X$ as some type of user information. To accommodate uncertainty in $X$, we define an internal profile as probabilities of ``community-$X$'', and an external profile as probabilities of ``community-community-$X$''. 
Here we focus on $X$ as \emph{content}, which is the primary user information in many social networks.
E.g., in Twitter, users write tweets and retweet from others; in DBLP, authors publish papers and cite papers from others. 
We call the probabilities of ``community-content'' as \emph{content profile} (i.e., what a community is about), and those of ``community-community-content'' as \emph{diffusion profile} (i.e., how a community diffuses certain content with another). 
Other types of $X$'s may exist in different networks, e.g., attributes in Facebook. Thus, ``community profile'' is a flexible concept. We leave other types of $X$'s as future work. 

\vspace{-2mm}
{\flushleft $\bullet$ {\it Is a community profile good}?} 
Due to network homophily, users in the same community tend to have similar behaviors. Thus, a community's profile should explain the common behaviors of its users. In other words, we do not see any ``community-content'' distribution as a good content profile; instead, only that well explaining the observations of user content as generated by the communities is a ``good'' one. Analogously, only the ``community-community-content'' distribution that well explains the observations of user-to-user content diffusion as generated by the communities is a ``good'' diffusion profile. This quality criterion will later guide us to estimate the profiles accurately. It is also a key to differentiating us from other work-- some prior attempt simply aggregates user information to output community properties (mostly internal ones) \cite{han:hal-01263772}, but it does not require such properties to best explain the  observations of user behaviors as generated by the communities through them. 

We consider ``community profiling'' as a new problem to solve due to three reasons. 
Firstly, community profiling is different from community detection, because detection focuses on getting community membership for each user, whereas profiling focuses on getting the ``community-content'' and ``community-community-content'' probabilities. 
Secondly, community profile has never been defined. Some recent work exploits rich user information \cite{conf/kdd/HanT15,RuanFP13,SachanDSXH14,SunAH12,XuKWCC12,YangML13,ZhouCY09} to improve community detection and output some user information aggregation as a by-product. But they neither define a community profile both internally and externally, nor try to identify new applications that community profiles enable. 
Finally, the difficulty of community profiling is often underestimated. As we shall discuss soon, due to the inter-dependency with community detection, the heterogeneity of social observations and the nonconformity of user behaviors, finding a good community profile is challenging. None of the existing work has ever identified and addressed such challenges (more discussions in Sect. \ref{sec:related_work}). 

Our goal is to infer content profile and diffusion profile for each community, and ultimately enable new applications. 
In Fig.~\ref{fig:framework}(a), we show the input for community profiling: a set of users, each of whom publishes documents; users are connected by friendship links, and interact with each other by diffusion links. 
E.g., in Twitter, each user posts tweets, users are connected by followership links, and they retweet each other to diffuse information. 
In Fig.~\ref{fig:framework}(b), for each community, we output: a content profile (e.g., community $c_1$ tends to publish topics $z_1$ and $z_2$) and a diffusion profile (e.g., $c_1$ tends to diffuse itself and $c_2$ on $z_1$). 
In Fig.~\ref{fig:framework}(c), we enable three new applications as follows (novelty to be discussed in Sect. \ref{sec:related_work}, applications to be concretized in Sect. \ref{sec:applications} and evaluated in Sect. \ref{sec:experiments}): 

\vspace{-2mm}
{\flushleft $\bullet$ {\it Community-aware diffusion}.} 
As community profiles aggregate user behaviors, we can use them to more robustly model the diffusion in a community level, rather than an individual level \cite{DuLBS14,LinWHY13,LucierOS15}. E.g., we can explain a retweet happens as one user's communities often retweet the other's on a certain topic. 
We acknowledge diffusion as a complex decision-- beyond community profiles, there are also nonconformity factors such as individual preference and topic popularity. This partially explains why community profiling is challenging-- we cannot account community profiles for all the diffusions; instead, we have to model different factors, to accurately estimate the profiles and the community-aware diffusion. 

\vspace{-2mm}
{\flushleft $\bullet$ {\it Profile-driven community ranking}.} We often need to target audiences for disseminating information in the networks. E.g., a company wants to target communities, which are most likely to retweet about its product, so as to launch a campaign. A funding agency wants to target communities, which actively cite papers about its grant theme on ``deep learning'', so as to disseminate the grant call. Since we have known what content each community is interested in and how it diffuses that content with others, we can rank the communities. Profile-driven community ranking is different from the traditional community recommendation, which often relies on only ``community-X'' properties and is unaware of diffusion \cite{Chen:2008:CCF,han:hal-01263772}. 

\vspace{-2mm}
{\flushleft $\bullet$ {\it Profile-driven community visualization}.} Holistic modeling leads to rich visualization-- we can now visualize not only how communities feature distinct contents (e.g., what an IT community tweets), but also how they interact (e.g., how an IT community retweets others) which is often overlooked before \cite{Cruz:2014:CDV,LinSCKSK09}.

We make two remarks about the above applications: 
1) we complete one task of community profiling to support multiple applications at a time, thus community profiling is only done once offline; 2) we build an interactive system\footnote{\url{http://sociallens.adsc.com.sg/}} for profile-driven community visualization and ranking, which  for the first time allows people to freely browse the communities by both content and diffusion \cite{CaiZZCH17}. 

The difficulty of community profiling is often largely underestimated; as we shall discuss next, there exist many challenges: 
\vspace{-2mm}
{\flushleft $\bullet$ {\it Inter-dependency with community detection}.} 
A straightforward approach of community profiling is to first detect communities and then aggregate each community's user observations as the profiles. However, because this approach does not try to ``best explain'' the user observations as generated by the communities through their profiles, it is often suboptimal. 
Take content profile as an example. 
Denote a user as $u$ and a community as $c$. 
For simplicity, we denote $c$'s content profile as $p(\text{content}|c)$ and the likelihood of $u$'s content as $p(\text{content}|u)$. 
To best explain the user content as generated by the communities through their content profiles, we effectively solve
\begin{equation} \label{eq.content_profile}
 \textstyle \max \prod_{u} p(\text{content}|u)  = \prod_{u} \sum_{c} p(\text{content}|c)p(c|u),
\end{equation}
where $p(c|u)$ is the probability of user $u$ assigned to community $c$. 
Ideally, to optimize Eq.~\ref{eq.content_profile}, we shall optimize both the profile $p(\text{content}|c)$'s and the community assignment $p(c|u)$'s. But in the straightforward approach, the detection first fixes the $p(c|u)$'s, then the best result this aggregation can return is the $p(\text{content}|c)$'s that maximize Eq.~\ref{eq.content_profile}. It is clear that, the maximal likelihood we get with fixed $p(c|u)$'s is suboptimal, unless the $p(c|u)$'s are ``perfect''. 
A perfect detection of $p(c|u)$'s also needs to maximize the likelihood in Eq.~\ref{eq.content_profile}, which depends on the profile $p(\text{content}|c)$'s. 
In all, content profiles and community detection are coupled. 
Similarly, we can show that diffusion profile and community detection are also coupled. 
Denote one more user as $v$, and one more community as $c'$. 
For simplicity, we denote $c$'s diffusion profile as $p(\text{diffusion}|\text{content}, c, c')$'s, each as a probability of having a diffusion  between $c$ and $c'$ about some content.
Then, to best explain the user-to-user diffusion as generated by the communities through their diffusion profiles, we effectively solve
\begin{align} 
     &\max \textstyle \prod_{(u,v)} p(\text{diffusion}|\text{content}, u, v)  \label{eq.diffusion_profile} \\
     &~~~~\textstyle= \prod_{(u,v)} \sum_{c}  \sum_{c'} p(\text{diffusion}| \text{content}, c, c') p(c|u) p(c'|v), \nonumber
\end{align}
where the product over $(u,v)$ is taken over all the user pairs having a diffusion link. 
To optimize the likelihood in Eq.~\ref{eq.diffusion_profile}, we shall optimize the diffusion profile $p(\text{diffusion}| \text{content}, c, c')$'s, as well as the community assignments $p(c|u)$'s and $p(c'|v)$'s.
\vspace{-2mm}
{\flushleft $\bullet$ {\it Heterogeneity of social observations}.} 
Social observations, especially the user links (i.e., friendship links and diffusion links), often carry different semantics; e.g., friendship links indicate user connections and diffusion links indicate user interactions. Traditionally, we often try to enforce user connections to be denser within each community than across communities \cite{HeLMCSY16,LeskovecLM10}. But in diffusion, the ``weak ties'' theory recognizes that the inter-community interactions may not be weak \cite{Granovetter83}. E.g., software engineering community cites more papers from machine learning community than itself on ``deep learning.'' This means we have to separate the modeling of user connection and user diffusion. Such user link heterogeneity is largely overlooked in the previous work \cite{SachanDSXH14,TangWL09}, thus how to model heterogeneous user links together remains unclear. 
\vspace{-2mm}
{\flushleft $\bullet$ {\it Nonconformity of user behaviors}.} 
User behaviors, especially their diffusion decisions, can happen for many reasons. Community-level conformity is just one reason, thus we have to consider other factors as well. E.g., some diffusion happens due to its topic (e.g., presidential election) being popular at the moment or its author (e.g., Lady Gaga) being preferred as a celebrity. Such topic popularity and user preference are the other two typical nonconformity factors for diffusion, and we must accommodate them. No prior work has explored both community factor and nonconformity factors \cite{HuYCX15,LucierOS15}, and it is not clear how to balance them in diffusion.  

Our technical novelty is identifying the above challenges and developing a unified Community Profiling and Detection (CPD) model (Sect. \ref{sec:cpd}) to address them accordingly. 
\vspace{-2mm}
{\flushleft $\bullet$} To model the \emph{inter-dependency with community detection}, we propose to take a novel profile-aware generative approach-- we realize the detection by latent membership variables and the profiling by latent community profile variables, which together generate the user friendship links, user content and user diffusion links in the network. Then we infer these latent variables by maximizing the likelihood. None of the existing work has taken a profile-aware generative approach-- they may use a generative model for community detection \cite{McAuleyL12,SachanCFS12,SunAH12}, but they never consider internal and external profiles together with detection.
\vspace{-2mm}
{\flushleft $\bullet$} To address the \emph{heterogeneity of social observations}, we propose to separate the generation of friendship links from latent community assignments and the generation of diffusion links from latent profiles. In particular, we require that two users are more likely to share a friendship link if they have similar community assignments. Thus maximizing the likelihood of observing the friendship links enforces  intra-community friendship links to be denser than inter-community ones. In contrast, we use the community diffusion profiles to generate the diffusion links, but we do not require inter-community diffusion strengths to be always smaller than intra-community ones; instead, the diffusion profiles are freely learned in maximizing the likelihood of the diffusion link observations. 
\vspace{-2mm}
{\flushleft $\bullet$} To accommodate the \emph{nonconformity of user behaviors}, we propose to define the generative probability of observing a diffusion link as a logistic function over multiple factors, including the topic-aware community diffusion profiles, the time-sensitive topic popularities and the individual user preferences. By maximizing the likelihood of diffusion link observations, we learn the diffusion profiles, as well as the weights to combine these different factors. 

Finally, we design a scalable inference algorithm for CPD (Sect. \ref{sec:inference}). 
As shown later, our inference algorithm scales linearly to the data set size. 
We further parallelize our inference algorithm, by taking the data skewness into account.

We summarize our contributions as follows:

\vspace{-2mm}
{\flushleft $\bullet$ We identify a \emph{new problem of community profiling}, which together with detection enables a holistic modeling of communities.} 

\vspace{-2mm}
{\flushleft $\bullet$ We identify \emph{three unique challenges} and design \emph{a novel CPD model} for joint community profiling and detection.} 

\vspace{-2mm}
{\flushleft $\bullet$ We develop \emph{a scalable inference algorithm} for CPD, and we further parallelize it by taking the data skewness into account.} 

\vspace{-2mm}
{\flushleft $\bullet$ We perform \emph{extensive experiments} to evaluate CPD over large-scale data sets, and show both its effectiveness and scalability.}

\section{Related Work} \label{sec:related_work} 
In this section, we review the related work on community detection and relevant applications, and distinguish the differences between existing work and our community profiling model. We further organize such differences in Table \ref{tab:comparedmethods}.

\begin{sidewaystable}
\centering
\vspace{0.2cm}
\tabcolsep=1mm
\begin{small}
\begin{tabular} { |c|c|c|c|c|c|c|c|c|c|c|c|c|} \hline
  & \multicolumn{5}{|c|}{\textbf{Data}}& \multicolumn{3}{|c|}{\textbf{Diffusion factors}}& \multicolumn{4}{|c|}{\textbf{Tasks}}  \\ \cline{2-13}
  \textbf{Methods} &text&attribute&node feature &friend. link & diff. link &individual &community &topic &topic extract. &community detect. & diffusion pred. & community profile\\ 
  \hline
 MaxFlow \cite{FlakeLG00}&&&&$\bullet$&&&&&&$\bullet$&&\\\hline
 SN-LDA \cite{SachanDSXH14} &$\bullet$&&&$\bullet$&&&&&$\bullet$&$\bullet$&&\\ \hline
 CODICIL \cite{RuanFP13}&$\bullet$&&&$\bullet$&&&&&&$\bullet$&&\\ \hline
 SocialCircle \cite{McAuleyL12} &&$\bullet$&&$\bullet$&&&&&&$\bullet$&&\\ \hline
 CESNA \cite{YangML13} &&$\bullet$&&$\bullet$&&&&&&$\bullet$&&\\ \hline
 BAGC \cite{XuKWCC12} &&$\bullet$&&$\bullet$&&&&&&$\bullet$&&\\ \hline
 SA-Cluster \cite{ZhouCY09} &&$\bullet$&&$\bullet$&&&&&&$\bullet$&&\\ \hline
 MetaFac \cite{LinSCKSK09} &$\bullet$&&user actions&$\bullet$&&&&&&$\bullet$&&\\ \hline
 PMM \cite{TangWL09} &&&interaction&$\bullet$&&&&&&$\bullet$&&\\ \hline
 TURCM \cite{SachanCFS12} &$\bullet$&&interactions&$\bullet$&&&&&$\bullet$&$\bullet$&&\\ \hline
 GenClus \cite{SunAH12} &&$\bullet$&&$\bullet$&&&&&&$\bullet$&&\\ \hline
 GF \cite{BasuRoy:2015:GRG} &&&user-item pairs&&&&&&&$\bullet$&&$\bullet$\\ \hline
 CFF \cite{Ntoutsi:2012:FGR} &&&user-item pairs&&&&&&&$\bullet$&&$\bullet$\\ \hline
 Influlearner \cite{DuLBS14} &$\bullet$&&&$\bullet$&$\bullet$&$\bullet$&&&&&$\bullet$&\\ \hline
 LADP \cite{LinWHY13} &$\bullet$&&&$\bullet$&$\bullet$&$\bullet$&&&&&$\bullet$&\\ \hline
   TopicInfluence \cite{LiuTHJY10} &$\bullet$&&&$\bullet$&$\bullet$&$\bullet$&&$\bullet$&$\bullet$&&$\bullet$&\\ \hline
   INFEST \cite{LucierOS15} &&&&$\bullet$&$\bullet$&$\bullet$&&&&&$\bullet$&\\ \hline
   topcgo \cite{EftekharGK13} &&&&$\bullet$&$\bullet$&$\bullet$&$\bullet$&&&&$\bullet$&\\ \hline
    BlackHole \cite{DBLP:conf/icde/LimKL16}&&&&$\bullet$&&&&&&$\bullet$&&\\\hline
    HAM \cite{HeLMCSY16}&&&&$\bullet$&&&&&&$\bullet$&&\\\hline
 PMTLM \cite{Zhu:2013}&$\bullet$&&&&$\bullet$&&&$\bullet$&$\bullet$&$\bullet$&$\bullet$&\\ \hline
WTM \cite{WangWBCZCH13}&$\bullet$&&user features&$\bullet$&$\bullet$&$\bullet$&&&&&$\bullet$&\\ \hline
CRM \cite{conf/kdd/HanT15}&&$\bullet$&&&$\bullet$&$\bullet$&$\bullet$&&&$\bullet$&$\bullet$&\\ \hline
COLD \cite{HuYCX15}&$\bullet$&&&&$\bullet$&&$\bullet$&&$\bullet$&$\bullet$&$\bullet$&\\ \hline
CPD (ours)&$\bullet$&&&$\bullet$&$\bullet$&$\bullet$&$\bullet$&$\bullet$&$\bullet$&$\bullet$&$\bullet$&$\bullet$\\ \hline
\end{tabular}
\end{small}
\vspace{-0.2cm}
\caption{Comparison with the related work.}
\label{tab:comparedmethods}
\end{sidewaystable} 

\vspace{0.05in}\noindent\textbf{Community Detection}.
Detecting communities from various networks has been extensively studied in the last decade. There exist comprehensive surveys \cite{Xie:2013, LeskovecLM10, Wadhwa:2014} on community detection, which review different community detection methods in terms of detection algorithms, quality measures, benchmarks and so on. 

Conventionally, a community is defined as a group of nodes, in which intra-group connections are much denser than inter-group ones \cite{FlakeLG00,WangWYZ15}. The pioneer community detection studies aim to generate the community membership for each node purely based on the links amongst them \cite{LeskovecLM10, WangWYZ15}. The prevalence of social networks offers a rich collection of user links to use for community detection, such as the followership in Twitter \cite{SachanDSXH14}, Flickr \cite{RuanFP13} and Facebook/Google+ \cite{McAuleyL12,YangML13}, the co-authorship in DBLP \cite{XuKWCC12,ZhouCY09}, the email exchange \cite{SachanDSXH14}. However, most of these existing work only consider one single type of links. There are other different types of user links; e.g., users comment/reply other users in digg \cite{LinSCKSK09}, contact/co-contact/co-subscribe other users in YouTube \cite{TangWL09}, follow/reply/retweet other users in Twitter \cite{SachanCFS12}. But these different links were often modeled in the same way. So far as we know, none of the existing community  work considers the heterogeneity among user links (i.e., friendship links and diffusion links) as we do. 

Recent studies start to exploit the rich user information, such as content \cite{SachanDSXH14}, attribute \cite{SunAH12,XuKWCC12}, action \cite{LinSCKSK09,SachanCFS12}, to improve the detection. 
Consequently, in addition to community membership, they also occationally output some ``community-$X$'' associations, such as ``community-content'' \cite{SachanDSXH14}, ``community-attribute'' \cite{SunAH12,XuKWCC12} and ``community-action'' \cite{LinSCKSK09,SachanCFS12}. In our work, we simultaneously discover communities and characterize them with both internal and external profiles. Although some forms of internal community profiles may be obtained in some prior work (\cite{LinSCKSK09,SachanCFS12,SachanDSXH14,SunAH12,XuKWCC12}) as the by-products, the external profiles are greatly overlooked. 

There are some recent studies on aggregating each community's user preferences as some form of community profiles, so as to enable item recommendation to each community. Their work is different from ours in two aspects. On one hand, most of these community recommendation studies are given the communities as input \cite{han:hal-01263772,Ronen:2014:RSM}. Even though some of them did try to detect communities \cite{BasuRoy:2015:GRG,Ntoutsi:2012:FGR}, their definition of a community is a group of users who share similar preferences to a recommended item, which is not based on network links at all. In contrast, our community is a group of densely connected users, who share similar interests and diffusion behaviours. On the other hand, their community profile is obtained by aggregating the users' preferences, which is usually based on a \emph{least misery} or \emph{aggregate voting} approach. In contrast, we formalize the community profiles as the probabilities of ``community-$X$'' and probabilities of ``community-community-$X$''. Besides, we estimate these community profiles by a generative model together with community detection. 

\vspace{0.05in}\noindent\textbf{Community-aware Applications}.
The community profiles deepen our understanding of the detected communities and thus benefit a lot of community-level applications. Here we review the related work to our three example applications, including community ranking, community diffusion and community visualization. Firstly, for community ranking, most of existing studies \cite{han:hal-01263772, Chen:2008:CCF} rank communities based on users' interests on them, i.e., to find the favourite communities for users. Moreover, the communities to be ranked are often already predefined over the networks. In our work, the communities are not provided as the input, and our focus is to rank communities by both their internal content profiles and external diffusion profiles together. This will help the company/author to choose the promising community to promote their products/papers as much as possible.  
Secondly, for community diffusion, in contrast to our community-level diffusion modelling, most diffusion models are at the individual level \cite{DuLBS14,LinWHY13,LucierOS15}. Recently, there are some studies that consider diffusion at the community level, but either the communities are predefined \cite{EftekharGK13} or the topic-awareness is overlooked \cite{conf/kdd/HanT15,HuYCX15}. Besides, unlike our modeling of various diffusion factors together, individual factor is missing in \cite{HuYCX15} and topic popularity factor is missing in \cite{conf/kdd/HanT15}. 
Last, but not the least, for community visualization, although a lot of efforts have been devoted to community detection, only a few of them further visualize the results to facilitate the deep analysis and semantic interpretation. In \cite{Cruz:2014:CDV}, the authors propose a community detection and visualization model, which differentiates the inner nodes and the border nodes for visualizing the interactions between communities. While their objective is to design a layout algorithm for clearly displaying the communities and their interactions, we focus on demonstrating the topic-aware user interaction strengths among the communities.

\section{Joint Profiling and Detection} \label{sec:cpd}

In the following, we first define some key notions; then we formulate the joint community profiling and detection problem. Table \ref{tab:notation} summarizes the notations used in this paper.  

\begin{table}[t]
	\centering  
	\tabcolsep=0.1cm
	\begin{small}
		\begin{tabular}{l l}
			\hline
			{\bfseries Notations}& {\bfseries  Description}        \\ \hline
			$|U|$, $|W|$        	& The number of users and words   	  \\ 
			$|C|$, $|Z|$	&The number of communities and topics \\
			$|F|$, $|E|$ & The number of friendship links and diffusion links\\			
			$d_{ui}$	& The $i$-th document published by user $u$	\\
			$D_u$ & The set of documents published by user $u$ \\ 
			$W_{ui}$ & The set of words in document $d_{ui}$\\
			$w_{uik}$	& The $k$-th word in document $d_{ui}$ 	\\
			$c_{ui}$, $z_{ui}$			& The community assignment and topic assignment for $d_{ui}$		\\
			$E_{ij}^t$			& A diffusion link from document $i$ to document $j$ at time $t$\\
			$F_{uv}$	& A friendship link from user $u$ to user $v$	\\
			$\boldpi_u$ 	& Multinomial distribution over communities specific to user $u$	\\
			$\boldtheta_c$	& Multinomial distribution over topics specific to community $c$	\\
			${\boldphi_z}$		& Multinomial distribution over words specific to topic $z$	\\
			$\eta_{c,c'z}$		& Probability of community $c$ diffusing community $c'$ on topic $z$	\\
			$\boldnu$  &The parameters for modeling individual diffusion preference \\
			$\alpha$, $\beta$, $\rho$	& Dirichlet priors \\
			\hline
		\end{tabular}
	\end{small} 
	\vspace{-0.2cm}
	\caption{Notations} 
	\label{tab:notation}  
\end{table}

\begin{definition} 
A \textbf{social graph} is $\mathcal{G}=(U,$ $D,F,E)$, where $u \in U$ is a user and $d \in D$ is a user published document. There are two types of links in $\mathcal{G}$. $F_{uv} \in F$ is a \emph{friendship link} from user $u$ to user $v$; $E_{ij} \in E$ is a \emph{diffusion link} from document $i$ to document $j$. Both types of links are directed. 
\end{definition} 
For a Twitter network, $D_u \subset D$ is the set of tweets posted by user $u$; $F_{uv}$ represents that user $u$ follows user $v$; $E_{ij}$ represents that tweet $i$ is a retweet of tweet $j$. 
For a DBLP network, $D_u$ is the set of papers published by author $u$; $F_{uv}$ represents that author $u$ co-authors with author $v$; $E_{ij}$ represents that paper $i$ cites paper $j$. 

To enable content modeling, we first define topic. 
\begin{definition} 
A \textbf{topic} $z \in \{1, \dots, |Z|\}$ is a $|W|$-dimensional multinomial distribution $\boldphi_z$ over words, where each dimension $\phi_{z,w}$ is the probability of a word $w \in \{1, \dots, |W|\}$ belonging to $z$.
\end{definition}

Then, we define the community membership, as well as our community content profile and diffusion profile. 
\begin{definition} 
A user $u$'s \textbf{community membership} is a $|C|$-dimensional multinomial distribution $\boldpi_u$, where each dimension $\pi_{u,c}$ is the probability of $u$ belonging to community $c$, $\forall c \in \{1, \dots, |C|\}$.
\end{definition}

\begin{definition} 
The \textbf{content profile} of community $c$ is a $|Z|$-dimensional multinomial distribution $\boldtheta_c$ over topics, where each dimension $\theta_{c,z}$ is the probability of $c$ discussing topic $z$.
\end{definition}

\begin{definition} 
The \textbf{diffusion profile} of community $c$ is a $|C|\times|Z|$-dimensional matrix $\boldeta_{c}$, where each entry $\eta_{c,c'z}$ is the probability of $c$ diffusing another community $c'$ on topic $z$.
\end{definition}
Take community $c_1$ in Fig.~\ref{fig:framework} as an example. As $c_1$'s users publish more content on $z_1$ and $z_2$, the resulting $\theta_{c_1,z_1}$ and $\theta_{c_1,z_2}$ are bigger. 
Besides, as $c_1$'s users often retweet/cite themselves on $z_1$, the resulting $\eta_{c_1,c_1z_1}$ is big. 
As motivated in Sect. \ref{sec:introduction}, we formalize a joint profiling and detection problem to solve in this paper.
\begin{problem} \label{problem} 
Given a social graph $\mathcal{G}=(U,D,F,E)$, the task of \textbf{joint community profiling and detection} is to infer: 1) each user $u$'s community membership $\boldpi_u$, $\forall u \in U$; 2) each community $c$'s content profile $\boldtheta_c$ and diffusion profile $\boldeta_{c}$, $\forall c \in \{1, \dots, |C|\}$. 
\end{problem}

\subsection{Model Design} \label{sec:modeldesign}

Next, we concretize our model design w.r.t. the three technical challenges for community profiling as discussed in Sect. \ref{sec:introduction}. We will later evaluate how well we address each challenge in Sect. \ref{exp:threeinsights}. 

\begin{figure}[t]
  \centering
    \includegraphics[width=\linewidth]{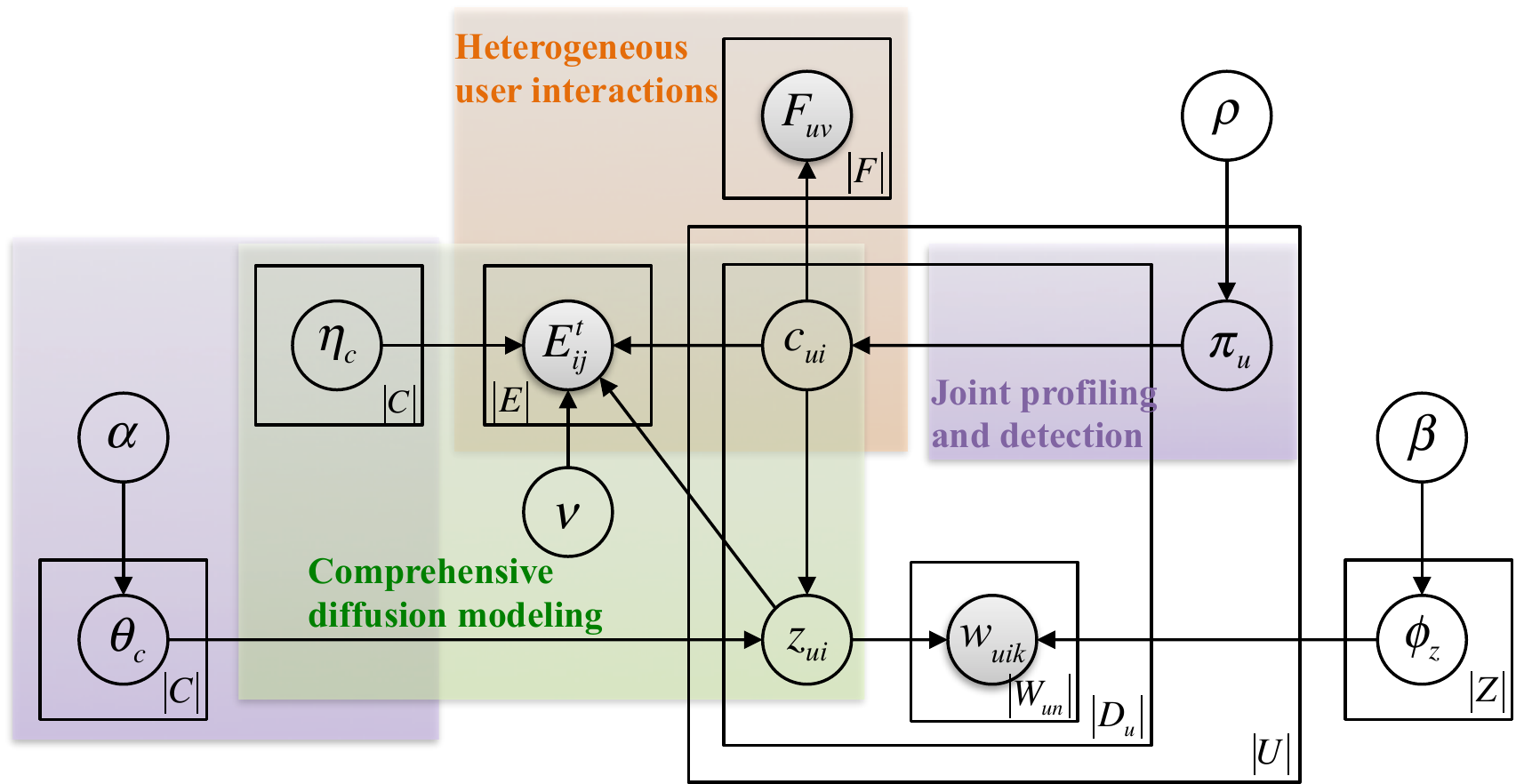}
    \vspace{-3mm}
  \caption{Graphical model of CPD.}
  \label{fig:cpd}
\end{figure}

\vspace{0.05in}\noindent\textbf{Profile-aware generative model}. 
Community detection aims to infer a community membership assignment $\boldpi_u$ for each user $u$ based on the friendship links $F_{uv}$'s. 
Community profiling aims to infer a content profile $\boldtheta_c$ and a diffusion profile $\boldeta_c$ for each community $c$ based on its member users' published content $D_u$'s and diffusion links $E_{ij}^t$'s. 
We can reinforce profiling and detection, by letting them leverage each other's data. 
As a result, we wish to infer a set of community-level latent variables, including $\boldpi_u$'s, $\boldtheta_c$'s and $\boldeta_c$'s, together from all the observations $(D, F, E)$. 

Since joint profiling and detection is an unsupervised task, we adopt a generative framework for our CPD model. 
We design CPD as a graphical model in Fig.~\ref{fig:cpd}, where we use communities to explain all the user observations on the network. 
Firstly, we consider a user $u$ to publish a document $d_{ui}$ of topic $z$, due to her community assignment $c_{ui}$ and the community content profile $\boldtheta_{c_{ui}}$. 
E.g., an author publishes a paper on deep learning, because she is from the machine learning (ML) community, which studies deep learning. 
As we deal with short documents (e.g., tweets in Twitter and paper titles in DBLP) and a short document is likely to be about one single topic \cite{SachanCFS12,HuYCX15}, we assign one single topic to each document in our model.
Secondly, we consider a user $u$ to publish a document $d_{ui}$ of topic $z$, which diffuses another user $v$'s document $d_{vj}$, due to both users' community assignments $c_{ui}$ and $c_{vj}$, as well as the community diffusion profile $\eta_{c_{ui},c_{vj}z}$. 
E.g., an author $u$ publishes a paper on software repositories, and cites another author $v$'s paper on deep learning, because $u$ is from the software engineering (SE) community, $v$ is from the ML community, and SE community tends to cite papers on deep learning from the ML community. 
Finally, we consider a user $u$ to form a friendship link with a user $v$, due to their similar community memberships $\boldpi_u$ and $\boldpi_v$. 
E.g., an author $u$ is a co-author with another author $v$, because they are both from the ML community. 

\vspace{0.05in}
\noindent\textbf{Addressing data heterogeneity}.
We model the friendship links $F$ and the diffusion links $E$ differently. Conventionally, a good community needs to have low conductance, which means the friendship links should be denser inside a community than outside a community. 
Specifically, we define the probability of having a friendship link between two users $u$ and $v$ as a sigmoid function, parameterized by their community membership similarity:
\begin{equation} \label{equ:puv}
   P(F_{uv}=1) = \sigma(\hat{\boldpi}_u^T \hat{\boldpi}_v),
\end{equation}
where $\hat{\boldpi}_u = [{\hat \pi}_{u,1}, ..., {\hat \pi}_{u,|C|}]^T$ is an estimation of $\boldpi_u$ based on the aggregation of $u$'s community assignments. In othe words, we use $\hat{\boldpi}_u$ and $\hat{\boldpi}_v$, instead of $\boldpi_u$ and $\boldpi_v$, to generate the $F_{uv}$'s in Fig.~\ref{fig:cpd}. Such a design is motivated by \cite{ChangB10,ChenZXZ13} to simplify the inference. $\sigma(x)=1/(1+e^{-x})$ is a sigmoid function. 
The more similar $\hat{\boldpi}_u$ and $\hat{\boldpi}_v$ are, the more likely $F_{uv}$ exists. In other words, $F_{uv}$ is large if $u$ and $v$ are from the same communities. This naturally enforces denser friendship links within a community than across communities, thus leading to low conductance. 
In contrast with the friendship links, the inter-community diffusion is not necessarily ``weak'' \cite{Granovetter83}. In fact, the community-level diffusion strengths vary over topics, which breaks the assumption of having to maintain the low conductance within a community. We need to resort to a different modeling of diffusion links, as we discuss next.

\vspace{0.05in}
\noindent\textbf{Accommondating nonconformity}. 
Different factors can account for a diffusion decision. Take Twitter as an example; user $u$ is likely to retweet $v$'s tweet $d_{vj}$ as her $i$-th tweet $d_{ui}$ at time $t$ if: 1) the community-level diffusion strength between $c_{ui}$ (the community $u$ belongs to when she generates document $d_{ui}$) and $c_{vj}$ on topic $z_{vj}$ is strong; 2) the topic $z_{vj}$ of $d_{vj}$ is trending at time $t$; 3) $u$ has an individual preference to retweet from $v$. These factors show three typical perspectives to make a diffusion decision: community perspective (if a community is more likely to retweet another community), content perspective (if a topic is more popular at the time) and user perspective (if a user is more likely to retweet another user). Next, we characterize the three typical factors.

\vspace{-2mm}
{\flushleft $\bullet$ \it{Community diffusion preference}:}
we consider a user $u$ to diffuse another user $v$ on topic $z$, if the communities of $u$ and $v$ are both interested in $z$ and they often diffuse each other on $z$. Denote $s \in \{0, 1\}$ as an indicator for a  diffusion link in $E$ to happen. Then, the probability of having a diffusion $s=1$ from $u$ to $v$ on $z$ is
 \begin{align}
  p(s=1,z|u,v) 
  &\overset{1}{=} \sum_c \sum_{c'} p(s=1|c,c',z) p(z|c) p(z|c') p(c'|v)  \nonumber \\
  &\overset{2}{\propto} \sum\nolimits_c \sum\nolimits_{c'} \eta_{c,c'z} {\hat \theta}_{c,z}  {\hat \pi}_{u,c} {\hat \theta}_{c',z} {\hat \pi}_{v,c'}, \label{equ:cf}
 \end{align}
where at step 1 we expand $p(s=1, z|u,v)$ by introducing the community membership $p(c|u)$ and $p(c'|v)$, the communities' interests on the topic $p(z|c)$ and $p(z|c')$, as well as the topic-sensitive community diffusion probability $p(s=1|c,c',z)$. 
At step 2, we estimate $p(s=1|c,c',z)$ with $\eta_{c,c'z}$, the probability of $c$ retweeting/citing $c'$ on $z$. Besides, we estimate $p(c|u)$ with ${\hat \pi}_{u,c}$, which is the empirical probability of community $c$ being assigned to user $u$; similarly we estimate $p(c'|v)$ with ${\hat \pi}_{v,c'}$. Finally we estimate $p(z|c)$ with $\hat \theta_{c,z}$, which is the empirical probability of topic $z$ assigned to the documents from $c$; similarly we estimate $p(z|c')$ with ${\hat \theta}_{c',z}$. 
Denote $\hat{\boldtheta}_{\cdot,z} = [{\hat \theta}_{1,z}, ..., {\hat \theta}_{|C|,z}]^T$ and $\bar{\boldeta}  = vec([\boldeta_1, ..., \boldeta_{|C|}])$, where $vec(\mathbf{A})$ concatenates the row vectors in a matrix $\mathbf{A}$ to a vector. 
For a diffusion between $d_{ui}$ and $d_{vj}$, which shares the same topic $z$, we denote $\bar{\mathbf{c}}_{ij} = vec( (\hat{\boldpi}_u \hat{\boldpi}_v^T) \circ (\hat{\boldtheta}_{\cdot,z} \hat{\boldtheta}_{\cdot,z}^T ))$, where $\circ$ is an element-wise product. Then Eq. (\ref{equ:cf}) becomes $\bar{\mathbf{c}}_{ij}^T \bar{\boldeta}$.

\vspace{-2mm}
{\flushleft $\bullet$ \it{Topic popularity}:} 
we model the popularity of a topic at a specific timestamp $t$ as the count of topic $z$ at $t$, which is denoted as $n^t_z$.

\vspace{-2mm}
{\flushleft $\bullet$ \it{Individual preference}:} 
we model user $u$'s preference to diffuse information from user $v$ with a linear function $\boldnu^T \mathbf{f}_{uv}$, where $\boldnu$ is a parameter, $\mathbf{f}_{uv}$ is a feature vector for $u$ and $v$. Take Twitter as an example; we consider two features for $u$: 1) \emph{user popularity}, which is defined as the number of $u$'s followers divided by that of her followees $\frac{|{Followers(u)}|}{|{Followees(u)}|}$; 2) \emph{user activeness}, which is defined as the number of $u$'s retweets divided by that of her tweets $\frac{|{Retweets(u)}|}{|{Tweets(u)}|}$. We extract $v$'s features and concatenate them with $u$'s as $\mathbf{f}_{uv}$. 

In order to systematically combine the three diffusion factors, we introduce a sigmoid function to define the probability of document $d_{vj}$ diffusing document $d_{ui}$ of topic $z$ at timestamp $t$ as: 
\begin{equation} \label{equ:pijt}
  p(E_{ij}^t=1|u,v,z,t) = \sigma(\bar{\mathbf{c}}_{ij}^T \bar{\boldeta} + n_z^t + \boldnu^T \mathbf{f}_{uv}).
\end{equation} 
We learn the parameters $\bar{\boldeta}$ and $\boldnu$, so that we know how much each factor contributes in the diffusion.

\vspace{0.05in} \noindent \textbf{Generative process}.
We summarize the CPD model's generative process below. Denote $\mathbbm{1}_{\ell \times 1}$ as an all-one vector of length $\ell$. 
\vspace{-0.1cm} 
\begin{enumerate}[leftmargin=*]
\itemsep=0in
\item For each topic $z=1, \dots, |Z|$, draw its word distribution from a Dirichlet prior parameterized by $\beta$: $\boldphi_z |\beta \sim Dir(\beta \mathbbm{1}_{|W| \times 1} )$; 
\vspace{-0.1cm} 
\item For each community $c= 1, \dots, |C|$, draw its topic distribution from a Dirichlet prior parameterized by $\alpha$: $\boldtheta_c |\alpha \sim Dir(\alpha \mathbbm{1}_{|Z| \times 1} )$; 
\vspace{-0.4cm} 
\item For each user $u = 1, \dots, |U|$ 
	\vspace{-0.2cm} 
	\begin{enumerate}[leftmargin=*]
	\itemsep=0in
	\item Draw her community distribution $\boldpi_u |\rho \sim Dir(\rho \mathbbm{1}_{|C| \times 1})$;
	\vspace{-0.05cm} 
	\item For the $i$-th document $d_{ui}$ of user $u$ 
		\vspace{-0.05cm} 
		\begin{enumerate}[leftmargin=*]
		\item Draw a community assignment $c_{ui}|\boldpi \sim Multi(\boldpi_u )$, by $u$'s multinomial community distribution $\boldpi_u$;
		\vspace{-0.03cm} 
		\item Draw a topic $z_{ui}|\mathbf{c},\boldtheta \sim Multi(\boldtheta_{c_{ui}} )$, by $c_{ui}$'s multinomial topic distribution $\boldtheta_c$; 
		\vspace{-0.03cm} 
		\item Draw each word $w_{uik}|\mathbf{z},\boldphi \sim Multi(\boldphi_{z_{ui}} )$, $\forall k = 1, ..., |W_{ui}|$, by $z_{ui}$'s multinomial word distribution; 
		\end{enumerate}	 
	\item For each friendship link from user $u$ to user $v$, draw ${F_{uv}}|\mathbf{\boldpi} \sim Ber( \sigma(\hat{\boldpi}_u^T \hat{\boldpi}_v) )$ by a Bernolli distribution (Eq.~\ref{equ:puv}); 
	\vspace{-0.05cm} 
	\item For each diffusion link $E^t_{ij}$ from document $d_{ui}$ to document $d_{vj}$ at time $t$, draw $E^t_{ij}|C,\boldeta, Z,\boldnu,\mathbf{f} \sim Ber( \sigma(\bar{\mathbf{c}}_{ij}^T \bar{\boldeta} + n_z^t + \boldnu^T \mathbf{f}_{uv}) )$ by a Bernolli distribution (Eq.~\ref{equ:pijt});
	\end{enumerate} 
\end{enumerate}
In step 3.b.iii, since short text often has single topic \cite{SachanCFS12,HuYCX15}, we sample all words in $d_{ui}$ from the same topic-word distribution $\boldphi_z$.

\section{Scalable Model Inference} \label{sec:inference}

We develop a scalable inference algorithm for CPD. 
We aim to infer the topic assignment and community assignment latent variables $\{ \mathbf{Z}, \mathbf{C} \}$ from the observations $\{W, F, E\}$, where $W$ is the words in $D$. We use collapsed Gibbs sampling \cite{ChenZXZ13,conf/kdd/HanT15,SachanDSXH14} for the inference. We also estimate the variational parameters $\{\boldpi, \boldtheta, \boldphi\}$ and the model parameters $\{\boldnu, \boldeta\}$ by variational Expectation Maximization (EM) \cite{ChangB10,ChenZXZ13}. We later parallelize our inference algorithm. 

\subsection{Collapsed Gibbs Sampling} \label{sec:collapsed_gibbs_sampling}
To derive the Gibbs sampler, we start with computing the collapsed posterior distribution of our model:
\begin{equation} \label{equ:joint}
\begin{aligned}
  &p(W,F,E,C,Z,\mathbf{f},\boldnu, \bar{\boldeta} |\rho ,\alpha ,\beta) \\
  &= p(C|\rho )p(Z|C,\alpha )p(W|Z,\beta )p(F|C)p(E|C,\boldeta,Z,\boldnu,\mathbf{f}), \\ 
\end{aligned} 
\end{equation}
where $p(F|C)$ (abbreviated as $p(F)$ in the following) is the probability for the friendship links $F$ generated by the communities $C$; $p(E|C,\bar{\boldeta},Z,\boldnu,\mathbf{f})$ (abbreviated as $p(E)$ in the following) is the probability for the diffusion links $E$ generated by the communities $C$. We follow \cite{ChangB10} to model observed links only in Eq.~\ref{equ:joint}. Thus, we define $p(F) = \prod\nolimits_{(u,v) \in F} P(F_{uv}=1) $ and $p(E) = \prod\nolimits_{(i,j) \in E} p(E_{ij}^t=1) $, where $t$ is the timestamp of the diffusion link $(i,j)$. 

In the generative process of CPD model, we model both $P(F_{uv}=1)$ (step 3.c) and $p(E_{ij}^t=1)$  (steps 3.d) with sigmoid functions $\sigma(\cdot)$. 
Bayesian inference with sigmoid function is known as hard, because it is analytically inconvenient to construct a Gibbs sampler for the sigmoid function \cite{PolsonSW13}. We are motivated by the data augmentation approach \cite{BiKWC14,ChenZXZ13}, which introduces P\'olya-Gamma random variables to derive an exact mixture representation of the sigmoid function for easier inference. Hence we introduce two P\'olya-Gamma variables $\boldlambda$ and $\bolddelta$ as the augmented variables for $p(F)$ and $p(E)$ respectively. 
Formally, a random variable $x$ follows a P\'olya-Gamma distribution $x \sim PG(a,b)$ ($a > 0, b > 0$), if
\[
x \textstyle = \frac{1}{2 \pi ^2} \sum_{k = 1}^\infty \frac{g_k}{ (k - 1/2)^2 + b^2/(4 \pi ^2) },
\]
where $g_k \sim Gamma(a,1)$ is a Gamma random variable. It has been shown in \cite{PolsonSW13}, a logistic function can be represented as a mixture of Gaussians w.r.t. a P\'olya-Gamma distribution:
\begin{equation}\label{equ:mg}
  \frac{1}{1 + e^{ - w}} = \frac{1}{2} \int_0^\infty \psi(w,x) p(x |1,0)dx,
\end{equation}
where $\psi(w, x) = e^{\frac{w-xw^2}{2}}$ and $x \sim PG(1,0)$. 
Then, for $p(F_{uv} = 1)$ as defined in Eq.~\ref{equ:puv}, we can introduce a P\'olya-Gamma variable $\lambda_{uv} \sim PG(1,0)$, such that we get a joint probability 
\begin{equation}
  p(F_{uv}=1, \lambda_{uv}) = \frac{1}{2} \psi(\hat{\boldpi}_u^T \hat{\boldpi}_v, \lambda_{uv}) p(\lambda_{uv} |1,0).
\end{equation}
Similarly, for $p(E_{ij}^t = 1)$ as defined in Eq.~\ref{equ:pijt}, we can introduce a P\'olya-Gamma variables $\delta_{ij} \sim PG(1,0)$, such that we get
\begin{equation}
p({E_{ij}^t}=1, \delta_{ij}) = \frac{1}{2} \psi(\bar{\mathbf{c}}_{ij}^T \bar{\boldeta} + n_z^t + \boldnu^T \mathbf{f}_{uv}, \delta_{ij}) p(\delta_{ij} |1,0).
\end{equation}
Considering all the friendship links and diffusion links, we have 
\begin{align}
  p(F,\boldlambda) &= \prod\nolimits_{(u,v) \in F} { \psi(\hat{\boldpi}_u^T \hat{\boldpi}_v, \lambda_{uv}) p(\lambda _{uv}|1,0) }, \label{equ:ag} \\
  p(E,\bolddelta ) &= \prod_{(i,j) \in E} { \psi( \bar{\mathbf{c}}_{ij}^T \bar{\boldeta} + n_z^t + \boldnu^T \mathbf{f}_{uv}, \delta_{ij} ) p(\delta_{ij}|1,0) }. \label{equ:ae}
\end{align}

Next we infer $\mathbf{Z}$ and $\mathbf{C}$, together with $\boldlambda$ and $\bolddelta$. 
Specifically, augmented with two P\'olya-Gamma variables $\boldlambda$ and $\bolddelta$, the collapsed posterior distribution of our model becomes:
\begin{equation} \label{equ:augjoin}
\begin{aligned}
&p(W,F,E,C,Z, \mathbf{f}, \boldnu ,\bar{\boldeta} ,\boldlambda ,\bolddelta |\rho ,\alpha ,\beta )\\
&= p(C|\rho )p(Z|C,\alpha )p(W|Z,\beta )p(F, \boldlambda | C)p(E,\bolddelta |C,\bar{\boldeta},Z,\boldnu ,\mathbf{f})\\
&=\textstyle \int_{\boldpi}  {P(C|\boldpi )P(\boldpi |\rho )} d\boldpi 
      \cdot \int_{\boldtheta}  {p(Z|C,\boldtheta )P(\boldtheta |\alpha )d\boldtheta } \cdot \\
&~~~\textstyle \int_{\boldphi}  {P(W|Z,\boldphi )P(\boldphi |\beta )d\boldphi } \cdot p(F,\boldlambda ) \cdot p(E,\bolddelta) \\
&\textstyle = \prod\limits_{u = 1}^{|U|} {\frac{{\Delta (n_u^{c} + \rho )}}{{\Delta (\rho )}}} \cdot\prod\limits_{c = 1}^{|C|} {\frac{{\Delta (n_c^{z} + \alpha )}}{{\Delta (\alpha )}}} \cdot\prod\limits_{z = 1}^{|Z|} {\frac{{\Delta (n_z^{w} + \beta )}}{{\Delta (\beta )}}} \cdot p(F,\boldlambda )\cdot p(E,\bolddelta ),
\end{aligned}
\end{equation}
where $\Delta(\mathbf{x}) = \frac{\prod_{i=1}^{\text{dim}(\mathbf{x})}\Gamma(x_i)}{\Gamma(\sum_{i=1}^{\text{dim}(\mathbf{x})}x_i)}$.
Based on Eq.~\ref{equ:augjoin}, we can infer $\mathbf{Z}$, $\mathbf{C}$, $\boldlambda$ and $\bolddelta$ one by one as follows.

\vspace{-2mm}
{\flushleft $\bullet$ \textbf{For $\mathbf{Z}$}:} the probability of assigning topic $z$ to $d_{ui}$ is
\begin{align}
&p({z_{ui}} \textstyle = z|C,{Z_{\neg \{ ui\} }},W,F,E, \mathbf{f},\boldnu ,\bar{\boldeta},\boldlambda,\bolddelta ) \nonumber \\
&\textstyle = \frac{{p(W,F,E,C,Z,\mathbf{f},\boldnu ,\bar{\boldeta} ,\boldlambda ,\bolddelta |\rho ,\alpha ,\beta )}}{{p(W,F,E,c_{ui} = c, C_{\neg \{ui \}},{Z_{\neg \{ ui\} }},\mathbf{f},\boldnu ,\bar{\boldeta} ,\boldlambda ,\bolddelta , \rho, \alpha ,\beta )}}  \nonumber \\
&\textstyle \propto \prod_{c = 1}^{|C|} {\frac{{\Delta (n_c^z + \alpha )}}{{\Delta (n_{c{,\neg }\{ ui\} }^z + \alpha )}}}  \cdot \prod_{z = 1}^{|Z|} {\frac{{\Delta (n_z^w + \beta )}}{{\Delta (n_{z{,\neg }\{ ui\} }^w + \beta )}}} \cdot \nonumber \\
&~~~~~~  p(F|\boldlambda, C_{\neg \{ui \}}) \cdot p(E|\bolddelta, C_{\neg \{ui \}}, Z_{\neg \{ ui\}} )  \nonumber \\
&\textstyle = \frac{{n_{c{,\neg }\{ ui\} }^z + \alpha }}{{n_{c{,\neg }\{ ui\} }^{(\cdot)} + \left| Z \right| \alpha }} \cdot \frac{{\prod\nolimits_{w = 1}^{\left| W \right|} {\prod\nolimits_{i = 1}^{n_{ui}^w} {(n_{z{,\neg }\{ ui\} }^w + \beta  + i - 1)} } }}{{\prod\nolimits_{j = 1}^{n_{ui}^{(\cdot)}} {(n_{z{,\neg }\{ ui\} }^{(\cdot)} + \left| W \right| \beta  + j - 1)} }} \cdot  \label{equ:sample_z} \\
&\textstyle ~~~~~ \prod\nolimits_{v \in \Lambda _u} \psi(\hat{\boldpi}_u^T \hat{\boldpi}_v, \lambda_{uv} | C_{\neg \{ui \}}) \cdot \nonumber \\ 
&\textstyle ~~~~~ \prod\nolimits_{j \in \Lambda _i} \psi(  \bar{\mathbf{c}}_{ij}^T \bar{\boldeta} + n_z^t + \boldnu^T \mathbf{f}_{uv}, \delta_{ij} | C_{\neg \{ui \}}, Z_{\neg \{ ui\}}), \nonumber 
\end{align}
where $\Lambda _u= \{ v | (u,v) \in F ~\text{or}~ (v,u) \in F\}$ is user $u$'s neighbors in $F$. $\Lambda _i = \{ j | (i,j) \in E  ~\text{or}~ (j,i) \in E \}$ is document $i$'s neighbors in $E$. $n_{c,{}\neg \{ ui\}}^z$ and ${n_{c,{}\neg \{ ui\} }^{( \cdot )}}$ denote the number of times that topic $z$ is assigned to community $c$ and the number of times that any topic is assigned to $c$, excluding the current document $d_{ui}$. Similarly, ${n_{z,{}\neg \{ ui\} }^w}$ and ${n_{z,{}\neg \{ ui\} }^{( \cdot )}}$ are the number of times that word $w$ is assigned to topic $z$ and the number of times that any word is assigned to $z$, excluding $d_{ui}$. ${n_{ui}^w}$ and ${n_{ui}^{( \cdot )}}$ are the number of times that word $w$ occurs in the document $d_{ui}$ and the number of words in $d_{ui}$. 
Finally, $\psi(\hat{\boldpi}_u^T \hat{\boldpi}_v, \lambda_{uv} | C_{\neg \{ui \}})$ denotes estimating $\psi(\hat{\boldpi}_u^T \hat{\boldpi}_v, \lambda_{uv})$ based on $C_{\neg \{ui \}}$ instead of the whole $C$; 
$\psi(  \bar{\mathbf{c}}_{ij}^T \bar{\boldeta} + n_z^t + \boldnu^T \mathbf{f}_{uv}, \delta_{ij} | C_{\neg \{ui \}}, Z_{\neg \{ ui\}})$ denotes estimating $\psi(  \bar{\mathbf{c}}_{ij}^T \bar{\boldeta} + n_z^t + \boldnu^T \mathbf{f}_{uv}, \delta_{ij})$ based on $C_{\neg \{ui \}}$ and  $Z_{\neg \{ ui\}}$ instead of the whole $C$ and $Z$. 

\vspace{-2mm}
{\flushleft $\bullet$ \textbf{For $\mathbf{C}$}:} the probability of assigning community $c$ to $u$ at $d_{ui}$ is
\begin{align}
  &p({c_{ui}} \textstyle = c|{C_{\neg \{ ui\} }},Z,W,F,E,\mathbf{f},\boldnu ,\bar{\boldeta} ,\boldlambda ,\bolddelta) \nonumber \\
 &\textstyle = \frac{{p(W,F,E,C,Z,\mathbf{f},\boldnu ,\bar{\boldeta} ,\boldlambda ,\bolddelta |\rho ,\alpha ,\beta )}}{{p({C_{\neg \{ ui\} }},z_{ui} = z, {Z_{\neg \{ ui\} }},W,F,E,\mathbf{f},\boldnu ,\bar{\boldeta} ,\boldlambda ,\bolddelta |\rho ,\alpha ,\beta )}}  \nonumber \\
 &\textstyle \propto \prod_{u=1}^{|U|} \frac{{\Delta (n_u^c + \rho )}}{{\Delta (n_{u{,\neg }\{ ui\} }^c + \rho )}}  \prod_{c=1}^{|C|} \frac{{\Delta (n_c^z + \alpha )}}{{\Delta (n_{c{,\neg }\{ ui\} }^z + \alpha )}} \cdot \nonumber \\
&~~~~~~ p(F|\boldlambda, C_{\neg \{ui \}}) \cdot p(E|\bolddelta, C_{\neg \{ui \}}, Z_{\neg \{ ui\}} )  \label{equ:sample_c} \\
&\textstyle = \frac{ n_{u{,\neg }\{ ui\} }^c + \rho } { n_{u{,\neg }\{ ui\} }^{(\cdot)} + \left| C \right| \rho } \cdot
        \frac{ n_{c{,\neg }\{ ui\} }^z + \alpha } { n_{c{,\neg }\{ ui\} }^{(\cdot)} + \left| Z \right| \alpha } \cdot \prod\nolimits_{v \in \Lambda _u} \psi(\hat{\boldpi}_u^T \hat{\boldpi}_v, \lambda_{uv} | C_{\neg \{ui \}}) \nonumber \\ 
&\textstyle ~~~~~ \cdot \prod\nolimits_{j \in \Lambda _i} \psi(  \bar{\mathbf{c}}_{ij}^T \bar{\boldeta} + n_z^t + \boldnu^T \mathbf{f}_{uv}, \delta_{ij} | C_{\neg \{ui \}}, Z_{\neg \{ ui\}}), \nonumber 
\end{align}
where ${n_{u,{}\neg \{ ui\} }^c}$ and ${n_{u,{}\neg \{ ui\} }^{( \cdot )}}$ are the number of documents from user $u$ that are assigned to community $c$ and the number of documents from user $u$ excluding $d_{ui}$, respectively.  

\vspace{-2mm}
{\flushleft $\bullet$ \textbf{For $\boldlambda$}:} the conditional distribution of $\boldlambda$ is P\'olya-Gamma, i.e., 
\begin{equation} \label{equ:sample_lambda}
\begin{array}{l}
    \textstyle p({\lambda _{uv}}|W,F,E,C,Z,\mathbf{f},\boldnu ,\boldeta ,\bolddelta ) \\
    \textstyle \propto e^{ \frac{ - \lambda_{uv} ( \hat{\boldpi}_u^T \hat{\boldpi}_v )^2}{2} } p({\lambda _{uv}}|1,0) = PG(1, \hat{\boldpi}_u^T \hat{\boldpi}_v).
\end{array}
\end{equation}
We efficiently sample $\lambda_{uv}$ by an alternate exponentially tilted Jacobi distribution \cite{PolsonSW13}.

\vspace{-2mm}
{\flushleft $\bullet$ \textbf{For $\bolddelta$}:} the conditional distribution of $\bolddelta$ is also P\'olya-Gamma,
\begin{align} 
   &\textstyle p({\delta _{ij}}|W,F,E,C,Z,\mathbf{f},\boldnu ,\boldeta ,\bolddelta ) \label{equ:sample_delta} \nonumber \\
   &\textstyle \propto e^{ \frac{ - \delta _{ij} (\bar{\mathbf{c}}_{ij}^T \bar{\boldeta} + n_z^t + \boldnu^T \mathbf{f}_{uv})^2 }{2} } p(\delta _{ij}|1,0) \\
   &\textstyle = PG(1, \bar{\mathbf{c}}_{ij}^T \bar{\boldeta} + n_z^t + \boldnu^T \mathbf{f}_{uv} ). \nonumber
\end{align}

\subsection{Model Parameter Estimation}
We use variational EM to iteratively estimate the variational parameters $\{\boldpi, \boldtheta, \boldphi\}$ and the model parameters $\{ \boldnu, \boldeta \}$:
\begin{enumerate}
\itemsep=0in
\item (E-step) Use the samples of collapsed Gibbs sampling to estimate the parameters $\boldpi$, $\boldtheta$ and $\boldphi$, given $\boldnu$, $\boldeta$.
\item (M-step) Optimize $\boldnu$ and $\boldeta$ by maximizing Eq. (\ref{equ:joint}), given the parameters $\boldpi$, $\boldtheta$, $\boldphi$ estimated in the E-step.
\end{enumerate}
In the E-step, the Gibbs sampler iteratively draws samples of $\mathbf{Z}$, $\mathbf{C}$, $\boldlambda$ and $\bolddelta$ by Eqs. \ref{equ:sample_z}--\ref{equ:sample_delta}. Based on the samples, we estimate: ${\pi _{u,c}} = \frac{{n_u^c + \rho }}{{n_u^{( \cdot )} + \left| C \right| \rho }}$, ${\theta _{c,z}} = \frac{{n_c^z + \alpha }}{{n_c^{( \cdot )} + \left| Z \right| \alpha }}$ and ${\phi _{z,w}} = \frac{{n_z^w + \beta }}{{n_z^{( \cdot )} + \left| W \right| \beta }}$.
In the M-step, we first estimate $\eta_{c,c'z}$'s by aggregating the community and topic assignments w.r.t all the documents, based on the last iteration of sampling. Then we estimate $\boldnu$ by maximizing Eq.~\ref{equ:joint} with all other variables fixed-- this is essentially fitting a logistic regression function; to solve it, we randomly sample the same amount of non-observed diffusion links as negative instances for optimization. As $\alpha$ and $\rho$ are used to sample the $\boldpi_u$'s and $\boldtheta_c$'s, we follow the convention \cite{griffiths_steyvers04} to set their values as 50 divided by $\boldpi_u$'s dimension and $\boldtheta_c$'s dimension respectively, i.e., $\alpha = 50/|Z|$, $\rho = 50/|C|$. As $\beta$ is used to sample the word distribution $\boldphi_z$'s and the number of words is large, we follow \cite{griffiths_steyvers04} again to set $\beta = 0.1$. 

\subsection{Scalability} \label{subsec:modelscalability} 

We summarize our inference algorithm in Alg.~\ref{alg.inference}. 
In steps 3--10, we take an E-step for collapsed Gibbs sampling. 
In steps 11-14, we take an M-step for training the model parameters. 

\vspace{0.05in}
\noindent \textbf{Time complexity}.
In steps 4--6, as we compute the community assignments and topic assignments for each document of each user, it takes $O(|D|\times|C|+|W|\times|Z|)$. In steps 7--8, as we compute $\hat{\boldpi}_u^T \hat{\boldpi}_v$ for each friendship link, it takes $O(|C|\times|F|)$. In steps 9--10, as we compute $(\bar{\mathbf{c}}_{ij}^T \bar{\boldeta} + n_z^t + \boldnu^T \mathbf{f}_{uv})$ for each diffusion link, it takes $|C|^2\times|E|$. In steps 11--12, as we aggregate the community assignments and topic assignments for each diffusion link, it takes $O(|E|)$. In steps 13--14, as we compute gradients for $\nu$ over all the diffusion links, it takes $O(|E|\times T_2)$. In total, for $T_1$ iterations, the overall complexity is $O((|D|\times|C|+ |W|\times|Z| + |C|\times|F| + |C|^2\times|E| + |E| + |E|\times T_2)\times T_1)$. As we can see, Alg.~\ref{alg.inference}'s time complexity is linear to the data size (i.e., $|D|$, $|F|$, and $|E|$). 

\begin{algorithm}[t]
\caption{Scalable inference for CPD}\label{alg.inference}
\begin{algorithmic}[1]
    \REQUIRE Users $U$, docs $D$, friendship links $F$, diffusion links $E$;
   \ENSURE Topic assignments $Z$, community assignments $C$, model parameters $\boldnu$ and $\boldeta$;
   \vspace{1mm}
   \STATE Initialize $\boldnu$, $\boldeta$, $\alpha$, $\beta$, $\rho$;
    \FOR{iter = 1:$T_1$}
    \FOR{each user $u \in U$}
	\FOR{each document $d_{ui} \in D_u$}
	     \STATE Sample a topic label $z_{ui}$ according to  Eq.~\ref{equ:sample_z};
	     \STATE Sample a community label $c_{ui}$ according to Eq.~\ref{equ:sample_c};
	\ENDFOR
    \ENDFOR
    \FOR{each friendship link $(u,v) \in F$}
	\STATE Sample augmented variable ${\lambda_{uv}}$ according to Eq.~\ref{equ:sample_lambda};
    \ENDFOR
    \FOR{each diffusion link $(i,j) \in E$}
	\STATE Sample augmented variable $\delta_{ij}$ according to Eq.~\ref{equ:sample_delta};
    \ENDFOR
    \FOR{each diffusion link $(i,j) \in E$}
	\STATE Update $\eta_{c_{ui}, c_{vj} z_{ui}}$ by aggregating $c_{ui}'s, c_{vj}$'s and $z_{ui}$'s;
    \ENDFOR
    \FOR{subiter = 1:$T_2$}
	\STATE Gradient descent for $\boldnu$ over the diffusion links $E$.
    \ENDFOR 
    \ENDFOR
\end{algorithmic}
\end{algorithm}

\vspace{0.05in}
\noindent \textbf{Parallelization}.
We consider multithread parallelization of Alg.~\ref{alg.inference}. 
We leave multi-machine parallelization as future work. 
In our variational EM algorithm, we find the E-step takes much longer time than the M-step, because: 
1) the E-step's collapsed Gibbs sampling has to be done iteratively over all the observations, including documents (thus words), friendship links and diffusion links; 
2) the M-step's model parameter estimation is comparatively much easier, since optimizing $\boldnu$ is basically solving logistic regression on the diffusion links (and the same amount of negative links) and $\boldeta$ is done by simply aggregating the existing community and topic assignments. 
Thus, in this paper we focus on parallelizing the E-step. 

\vspace{-2mm}
{\flushleft $\bullet$ \it{Segmenting data to reduce inter-dependency}}. 
Recall in Sect.~\ref{sec:collapsed_gibbs_sampling}, the sampling requires computing: 1) a number of counters, including the community-topic counter ${n_{c}^{z }}$, the topic-word counter ${n_{z}^{w }}$, the user-community counter ${n_{u}^{c}}$; 2) a number of link probabilities, including the friendship one $\psi(\hat{\boldpi}_u^T \hat{\boldpi}_v, \lambda_{uv})$ and the diffusion one $\psi(  \bar{\mathbf{c}}_{ij}^T \bar{\boldeta} + n_z^t + \boldnu^T \mathbf{f}_{uv}, \delta_{ij} )$. 
Among these computations, both topic and community assignments are applied to documents (thus their users), the friendship link probability is applied to users, and the diffusion link probability is applied to two documents (thus their users). Therefore, except the word topic assignment, the vast majority of computations are done on users and documents. This motivates us to segment the data by users and documents, so that different threads can work on different data segments with little inter-dependency. 
It may be possible to take words into consideration for the data segment as well, but it is not obvious and we leave it for future work. 
Considering that a user often has many documents (especially in Twitter), we design two guidelines to segment the data by users and documents: 
1) we keep a user's documents in the same data segment, because otherwise there are likely many conflicting updates about the same user from multiple threads; 
2) we prefer keeping the same-topic documents in the same data segment, because it helps to reduce the conflicting updates about the same topic from multiple threads. 
Overall, we first run LDA \cite{BleiNJ03} on all the users' documents with $|Z|$ topics; then we partition the users into $|Z|$ segments, based on each user's most frequently assigned topic in her documents. In each segment, each user has her documents, related friendship links and diffusion links. 

\vspace{-2mm}
{\flushleft $\bullet$ \it{Distributing workload to avoid data skewness}}. 
We aim to distribute the $|Z|$ data segments to $M$ threads, such that the workload on each thread is balanced. Note that $M$ is set as the number of physical CPU cores in this work. 
Our approach is to first estimate the workload of each data segment, and then cast this segment allocation task as solving $M$ standard \emph{0-1 knapsack problems}\footnote{\small{\url{https://en.wikipedia.org/wiki/Knapsack_problem}}}. Denote the $i$-th data segment's workload as $o_i \in \mathbb{R}^+$, thus the workload for all the data segments is $O = \sum_{i=1}^M o_i$. Denote a binary indicator as $x_i \in \{0, 1\}$. Then for each thread, we solve 
\begin{equation} \label{eq.knapsack}
  \textstyle \max~ \sum_{i=1}^M o_i x_i, ~~~\text{s.t.}~~~ \sum_{i=1}^M o_i x_i \leq \frac{1}{M} O,
\end{equation}
which tries to find a subset of the data segments to have as close to $\frac{O}{M}$ workload as possible. 
One can fine tune the objective function of Eq.~\ref{eq.knapsack} in practice to best allocate the data segments for even workload among the threads. 
We estimate each workload $o_i$ as follows. 
First of all, we estimate the average processing time for each document and each link, based on a serial implementation of the sampling algorithm over all the data. 
Then, based on the number of documents and links a user has, we estimate the average workload of processing that user. 
Finally, we sum up the average workload of all the users in the $i$-th data segment as $o_i$. 

We evaluate our inference algorithm's efficiency in Sect. \ref{sec.scalability}.

\section{Applications} \label{sec:applications}

We concretize how to enable the following three community-aware applications based on five CPD outputs, including: 
1) the community assignment for users $\pi_{u,c}$'s; 2) the community content profile $\theta_{c,z}$'s; 3) the community diffusion profile $\eta_{c,c'z}$'s; 4) the topic assignment for words $\phi_{w,z}$'s; 5) the individual diffusion preference parameters $\boldnu$.

\vspace{0.05in}
\noindent \textbf{Community-aware diffusion}.
Given input of a document $d_{vj}$ published by user $v$, we output the probability that another user $u$ will publish a document $d_{ui}$ to retweet or cite $d_{vj}$ at timestamp $t$ as
\begin{equation} \label{eq.community_diffusion}
 \begin{aligned}
   &\textstyle p(E_{ij}^t = 1 | u, v, d_{vj}, t) \overset{1}{=} \sum_z p(E_{ij}^t = 1 | u, v, z, t) p(z | d_{vj}) \\
   &\textstyle \overset{2}{=} \sum_z \sigma ( \sum_c \sum_{c'} \pi_{u,c} \theta_{c,z} \eta_{c,c'z} \pi_{v,c'} \theta_{c',z} + n_z^t + \boldnu^T \mathbf{f}_{uv} ) p(z | d_{vj}).
  \end{aligned}
\end{equation}
where at step 1 we expand $p(E_{ij}^t = 1 | u, v, d_{vj}, t)$ by the topics of $d_{vj}$. 
At step 2, we plug in the definition of $p(E_{ij}^t = 1 | u, v, z, t)$ by Eq.~\ref{equ:pijt}. 
As we can see, Eq.~\ref{eq.community_diffusion} comprehensively models the diffusion by taking the community assignments $\boldpi$, the community profiles $\boldtheta$ and $\boldeta$, and the individual diffusion preference $\boldnu$ into account. 

\vspace{0.05in}
\noindent \textbf{Profile-driven community ranking}. 
Given input of a query $q \in W^k$ ($\forall k \geq 1$), we output the ranking of communities based on their probabilities to diffuse information about $q$. Denote the probability of a community $c$ to generate a diffusion link $s=1$ of query $q$ as
\begin{equation} \label{eq.community_ranking}
 \begin{aligned}
   &\textstyle p(s=1|c,q) \overset{1}{=} \sum_z \sum_{c'} p(s=1|c,c',z) p(z|q,c')p(c'|q) \\
   &\textstyle \overset{2}{\propto} \sum_z \sum_{c'} \eta_{c,c'z} p(z|q,c') 
   \overset{3}{\propto} \sum_z \sum_{c'} \eta_{c,c'z} \theta_{c'z} \prod_{w\in q} \phi_{z,w}, 
 \end{aligned}
\end{equation}
where at step 1 we expand $p(s=1|c,q)$ by the community diffusion profile $p(s=1|c,c',z)$, the topic assignment for $q$ in a community $p(z|q,c')$ and the probability that $q$ is from that community $p(c'|q)$. 
At step 2 we plug in the definition of $p(s=1|c,c',z) \propto \eta_{c,c'z}$ and consider $q$ can come from any community with $p(c'|q)$ uniformly. 
At step 3, we estimate the probability $p(z|q,c')$ in a similar way as Eq.~\ref{equ:sample_z}.
We skip the details but explain the rational of this estimation: $p(z|q,c')$ is proportional to the probability of community $c'$ generating topic $z$ (i.e., captured by $\theta_{c',z}$) and the probability of $q$ belonging to topic $z$ (i.e., captured by $\prod_{w\in q} \phi_{z,w}$).

\vspace{0.05in}
\noindent \textbf{Profile-driven community visualization}. 
We can visualize each community's content profile and its diffusion profile, as Fig.~\ref{fig:framework}(b) shows. 
In particular, we are interested in the diffusion visualization, as it is new. In our experiments, we visualize how a community interacts with the others in two typical settings: 1) \emph{diffusion on a specific topic}, where we use $\eta_{c,c'z}$ as the diffusion strength from $c$ to $c'$ under topic $z$; 2) \emph{diffusion with topic aggregation}, where we use $\sum_z \eta_{c,c'z}$ as the diffusion strength from $c$ to $c'$. 

\section{Experiments} \label{sec:experiments}
We test CPD with two large-scale real-world data sets. We design experiments to: 1) evaluate how well we address each challenge listed in Sect.~\ref{sec:introduction}; 2) evaluate CPD's performance, by comparing with the state-of-the-art baselines in different applications.

\subsection{Set Up} \label{sec:setup}
We do experiments on Linux computers equipped with Intel(R) 3.50GHz CPUs and 16GB RAMs. We do 10-fold cross validation and report average scores for all the quantitative results. We also report significant test results whenever necessary.

\vspace{0.05in}
\noindent \textbf{Data sets}. 
We use two public data sets: Twitter \cite{DBLP:conf/kdd/LiWDWC12} and DBLP \cite{Tang:12TKDE}. 
The Twitter data set was collected in May 2011. 
The DBLP data set contains the publications indexed by DBLP\footnote{\small{\url{http://dblp.uni-trier.de/}}} from 1936 to 2010. 
We pre-processed the tweets and the paper titles, by removing stop words, stemming and POS tagging\footnote{\small{\url{http://nlp.stanford.edu/software/tagger.shtml}}}. We only kept nouns, verbs and hashtags. 
After that, we remove the documents with less than two words, and then remove the users with no document. 
Table \ref{tab:dataset} summarizes the statistics of our data sets after pre-processing.

\begin{table}[t]
\tabcolsep=0.12cm
	\centering  
	\begin{small}
		\begin{tabular}{l l l l l l}
			\hline
			 & \#(user) & \#(friend. link) & \#(diff. link) & \#(doc.) & \#(word)   \\ \hline
			Twitter & 137,325 & 3,589,811 & 992,522 & 39,952,379 & 2,316,020 \\ \hline
			DBLP & 916,907 & 3,063,186 & 10,210,652 & 4,121,213 & 330,334 \\ \hline
		\end{tabular}
	\end{small} 
	\vspace{-3mm}
	\caption{Data set statistics.} 
	\label{tab:dataset}  
\end{table} 

\vspace{0.05in}
\noindent \textbf{Baselines}. 
We choose baselines based on the following guidelines: 
1) they are the state of the art to model heterogeneous user observations at the data level; 2) they model diffusion prediction at the task level; 3) preferably they model community. 
Finally, we choose four baselines below, and list our differences with them in Table \ref{tab:comparedmethods}. 

\begin{table}[t]
\centering
\tabcolsep=0.05cm
\begin{small}
\begin{tabular} { | c|c |c|c|c|c|c|c|c|c|c|} \hline
  & \multicolumn{3}{|c|}{\textbf{Data}}& \multicolumn{3}{|c|}{\textbf{Diffusion factors}}& \multicolumn{4}{|c|}{\textbf{Tasks}}  \\ \cline{2-11}
  \textbf{Methods} &text &friend& diff&indiv- &comm &topic&topic  &comm  & diff & comm\\ 
  &&links&links&idual&&&extract&detect&pred&profile\\
  \hline
PMTLM \cite{Zhu:2013}&$\bullet$&&$\bullet$&&&$\bullet$&$\bullet$&&$\bullet$&\\ \hline
WTM \cite{WangWBCZCH13}&$\bullet$&$\bullet$&$\bullet$&$\bullet$&&&&&$\bullet$&\\ \hline
CRM \cite{conf/kdd/HanT15}&&$\bullet$&$\bullet$&$\bullet$&$\bullet$&&&$\bullet$&$\bullet$&\\ \hline
COLD \cite{HuYCX15}&$\bullet$&&$\bullet$&&$\bullet$&&$\bullet$&$\bullet$&$\bullet$&\\ \hline
Ours&$\bullet$&$\bullet$&$\bullet$&$\bullet$&$\bullet$&$\bullet$&$\bullet$&$\bullet$&$\bullet$&$\bullet$\\ \hline
\end{tabular}
\end{small}
\vspace{-3mm}
\caption{Differences with baselines.}
\label{tab:comparedmethods}
\end{table}

\vspace{-2mm}
{\flushleft $\bullet$ \emph{Poisson Mixed-Topic Link Model (PMTLM)} \cite{Zhu:2013}.} 
 It models the document network and uses the document topic assignment to generate the links. 
We adapt PMTLM for community detection and friendship link prediction comparison, by aggregating the topic assignments of each user's documents as the community membership for that user. We also compare with PMTLM on diffusion prediction, as it also models the document links. 

\vspace{-2mm}
{\flushleft $\bullet$ \emph{Whom to Mention (WTM)} \cite{WangWBCZCH13}.} 
It models the user diffusion links with user content and friendship. It does not model community. We compare with WTM on diffusion prediction. 

\vspace{-2mm}
{\flushleft $\bullet$ \emph{Community Role Model (CRM)} \cite{conf/kdd/HanT15}.} 
It models friendship links and diffusion links based on the user's community assignment and role assignment together. We compare with CRM on community detection, friendship link prediction and diffusion prediction. 

\vspace{-2mm}
{\flushleft $\bullet$ \emph{COmmunity Level Diffusion (COLD)} \cite{HuYCX15}.} 
It models the content and diffusion links based on communities. Thus it is the closest work to ours. But it models neither friendship links in community detection, nor individual factor and topic factor in diffusion prediction. We compare with COLD on community detection, friendship link prediction and diffusion prediction. As COLD has community diffusion strength, we also compare it on community ranking.

\begin{figure*}[t]
\small
\begin{tabular}[t]{c} \hspace{-0.3cm}
    \subfigure[Community detect. (Twitter)]{\label{fig:twitterhjcon}
        \includegraphics[width=0.24\linewidth]{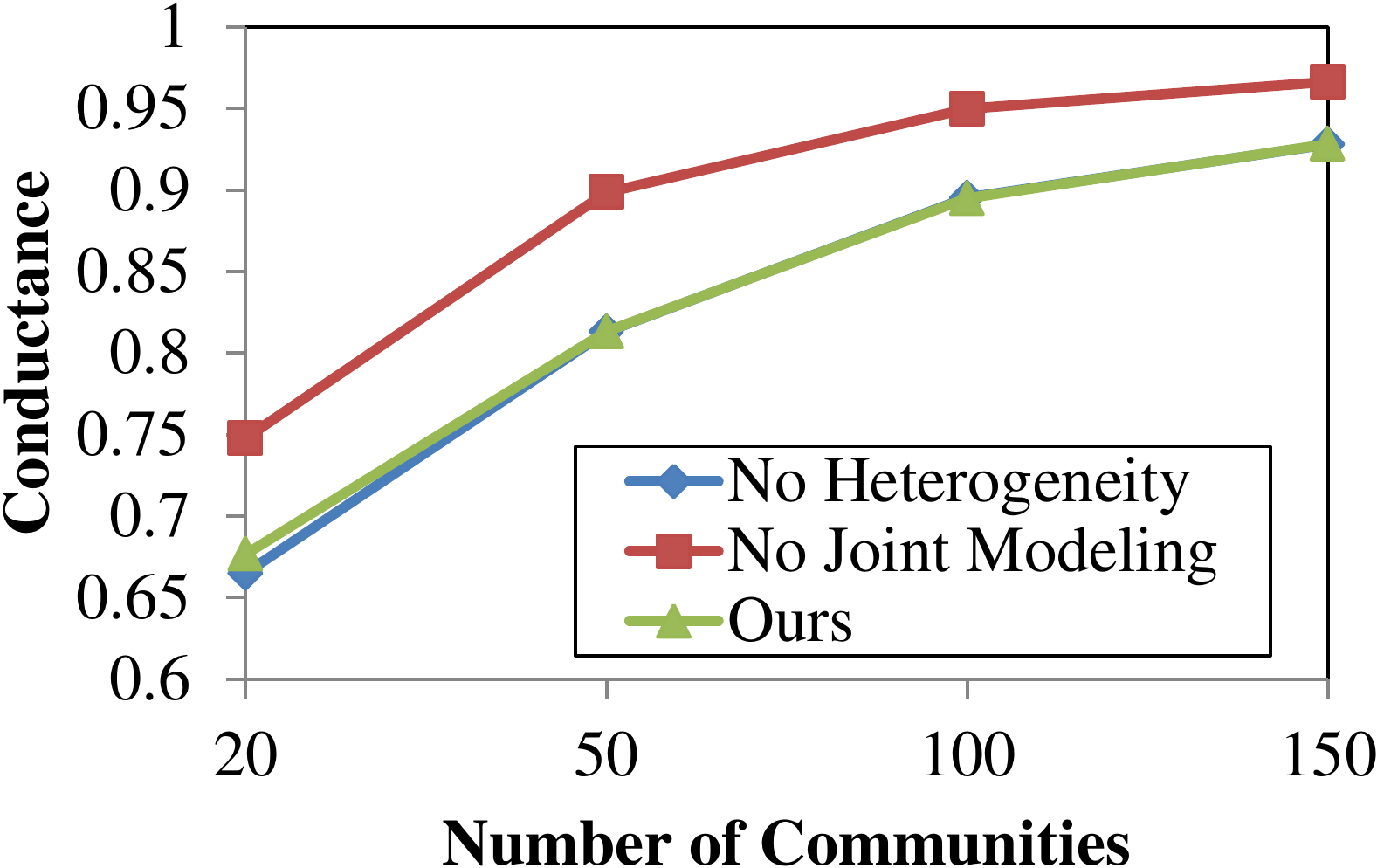}} 
    \subfigure[Friendship link pred. (Twitter)]{\label{fig:twitterhjlp}
        \includegraphics[width=0.24\linewidth]{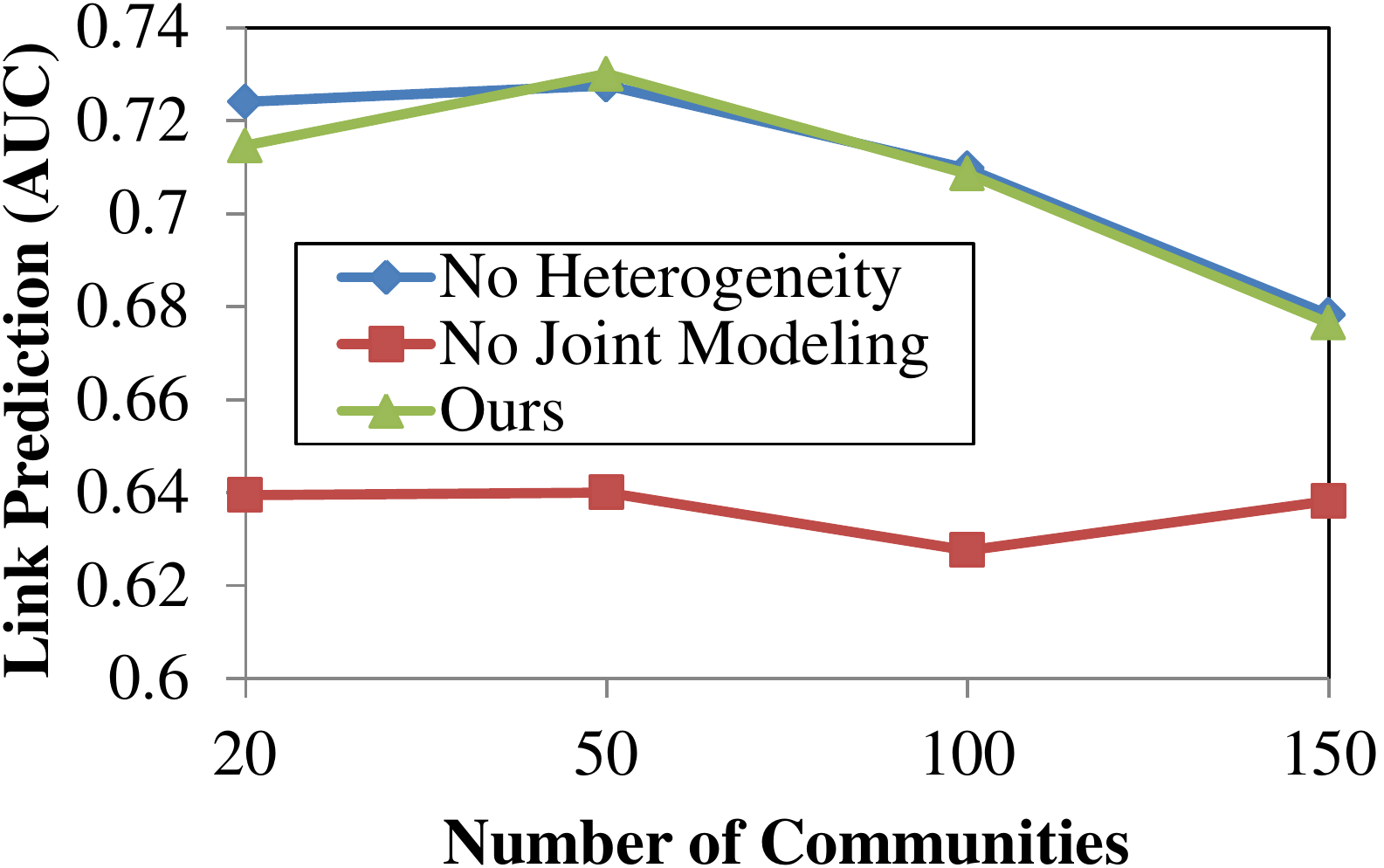}}
    \subfigure[Diffusion link pred. (Twitter)]{\label{fig:twitterhjdp}
        \includegraphics[width=0.24\linewidth]{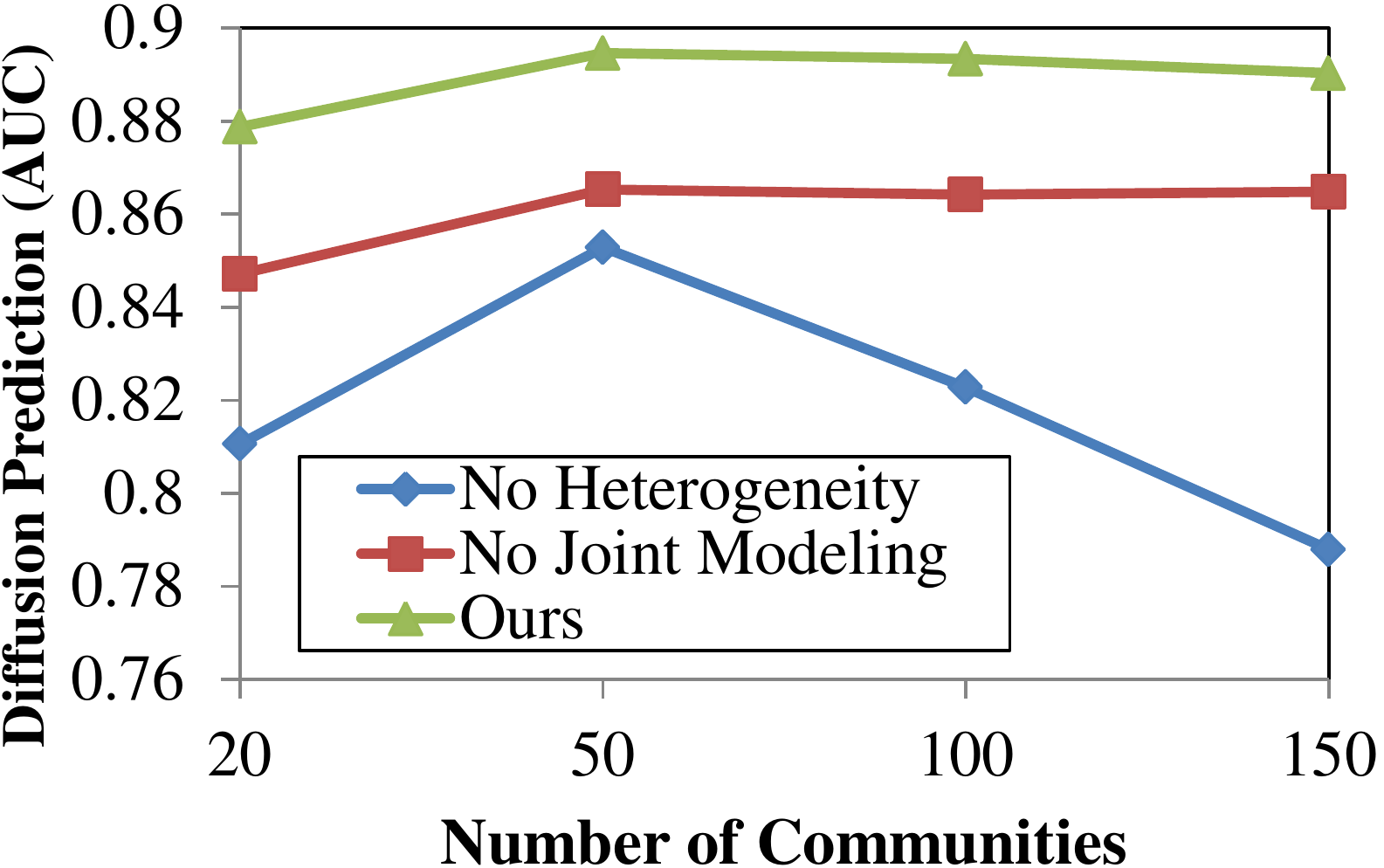}}
    \subfigure[Community detect. (DBLP)]{\label{fig:dblphjcon}
        \includegraphics[width=0.24\linewidth]{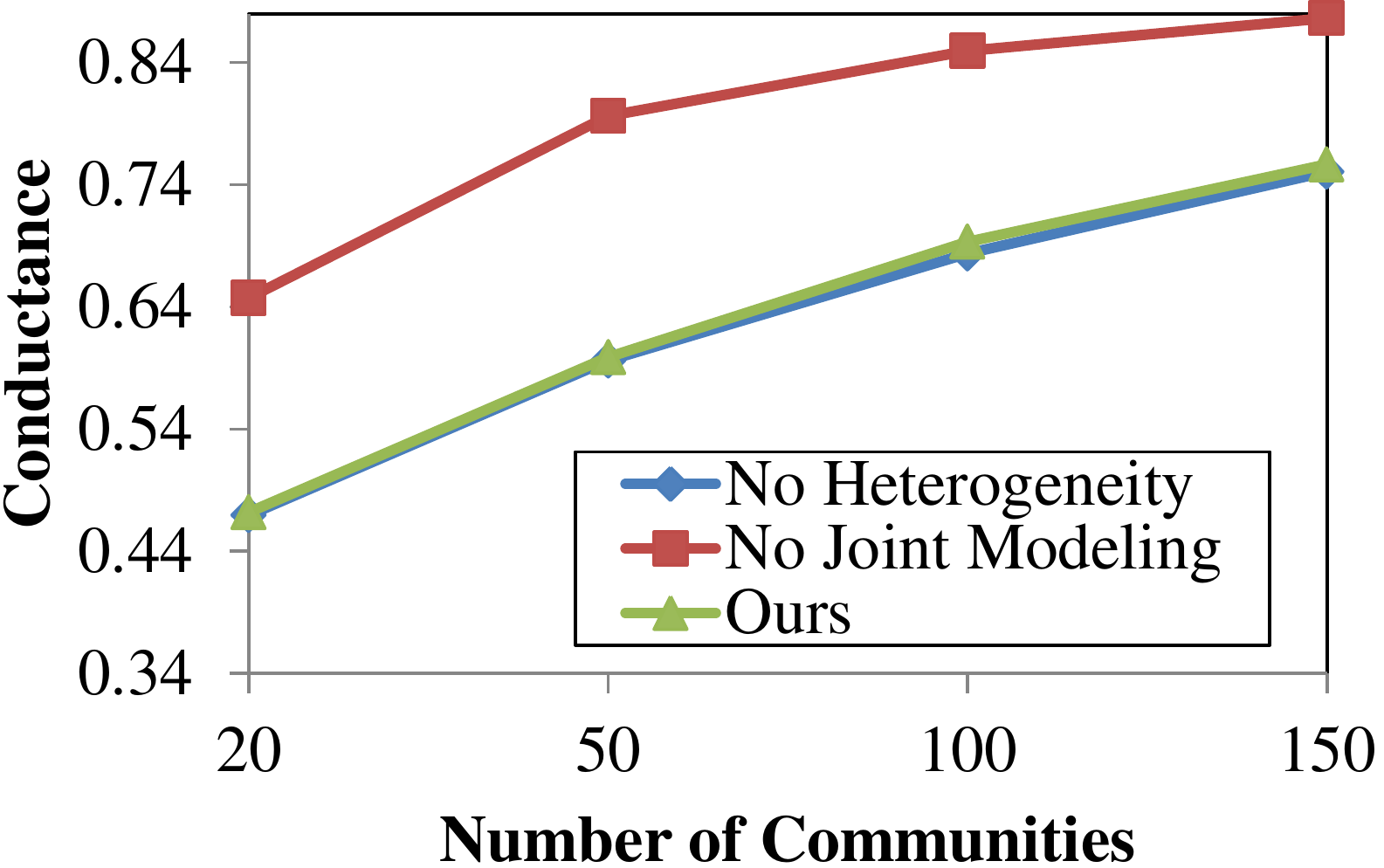}}\\ \hspace{-0.3cm}
    \subfigure[Friendship link pred. (DBLP)]{\label{fig:dblphjlp}
        \includegraphics[width=0.24\linewidth]{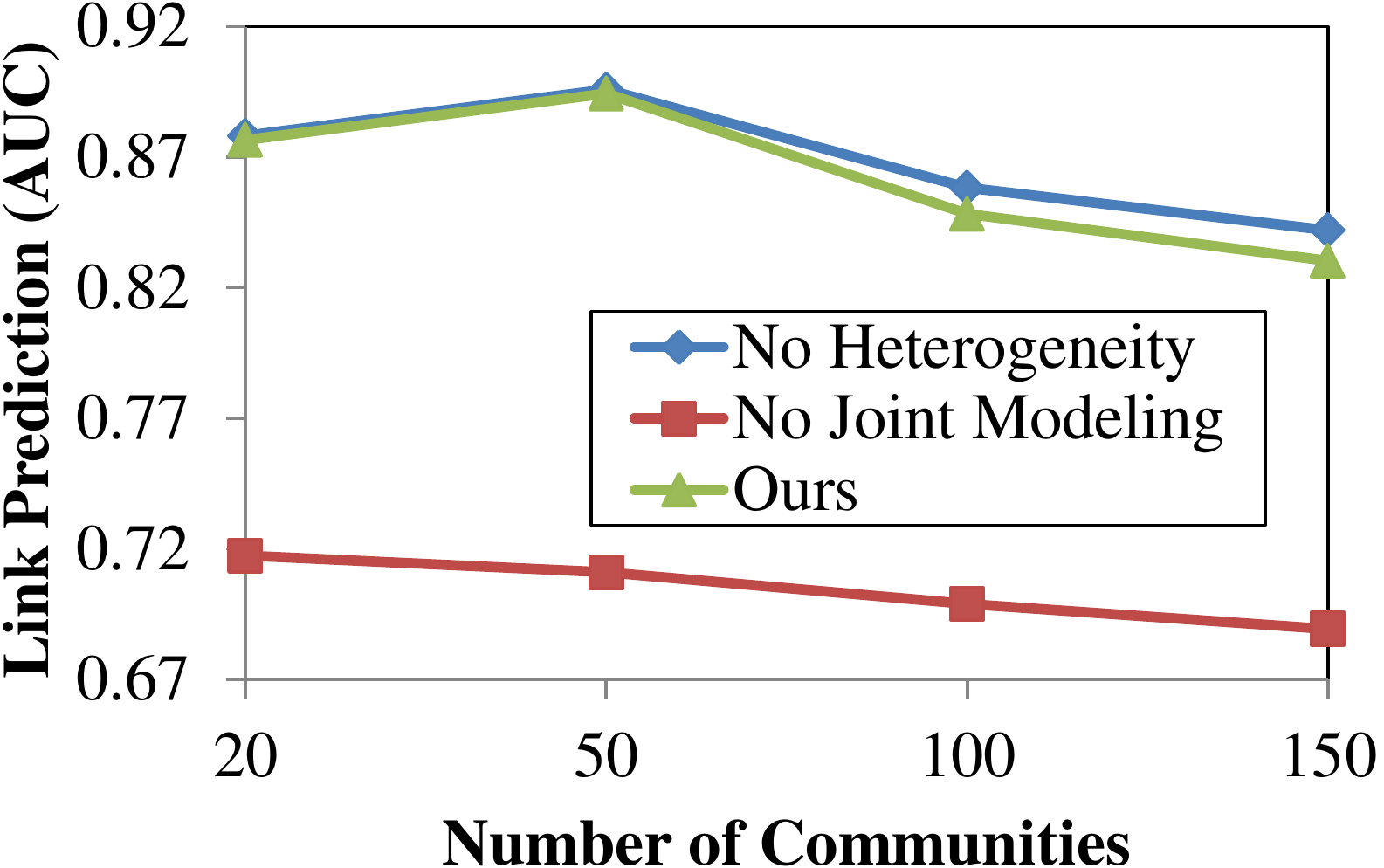}}
    \subfigure[Diffusion link pred. (DBLP)]{\label{fig:dblphjdp}
        \includegraphics[width=0.24\linewidth]{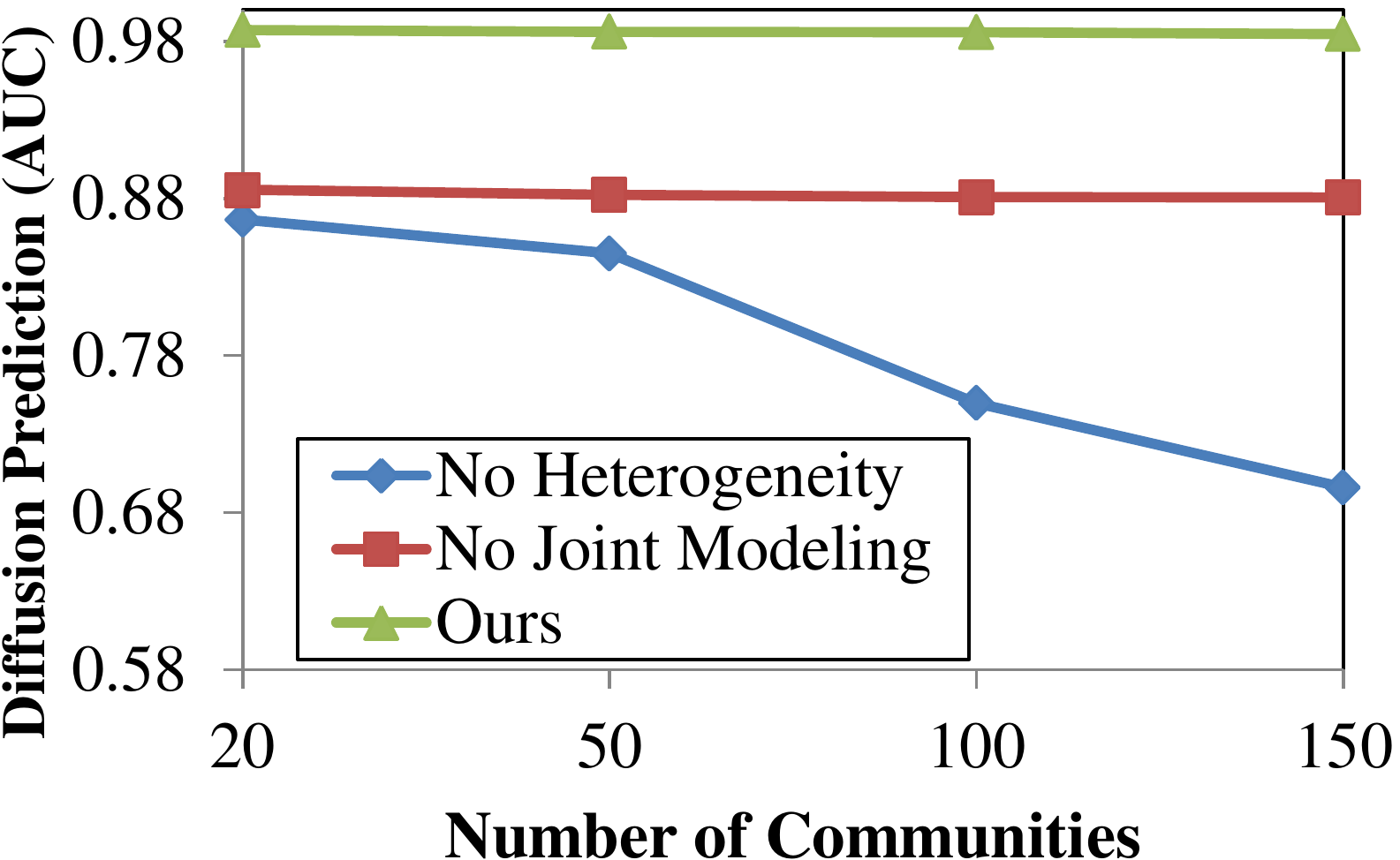}}     
    \subfigure[Diffusion link pred. (Twitter)]{\label{fig:twittercdm}
        \includegraphics[width=0.24\linewidth]{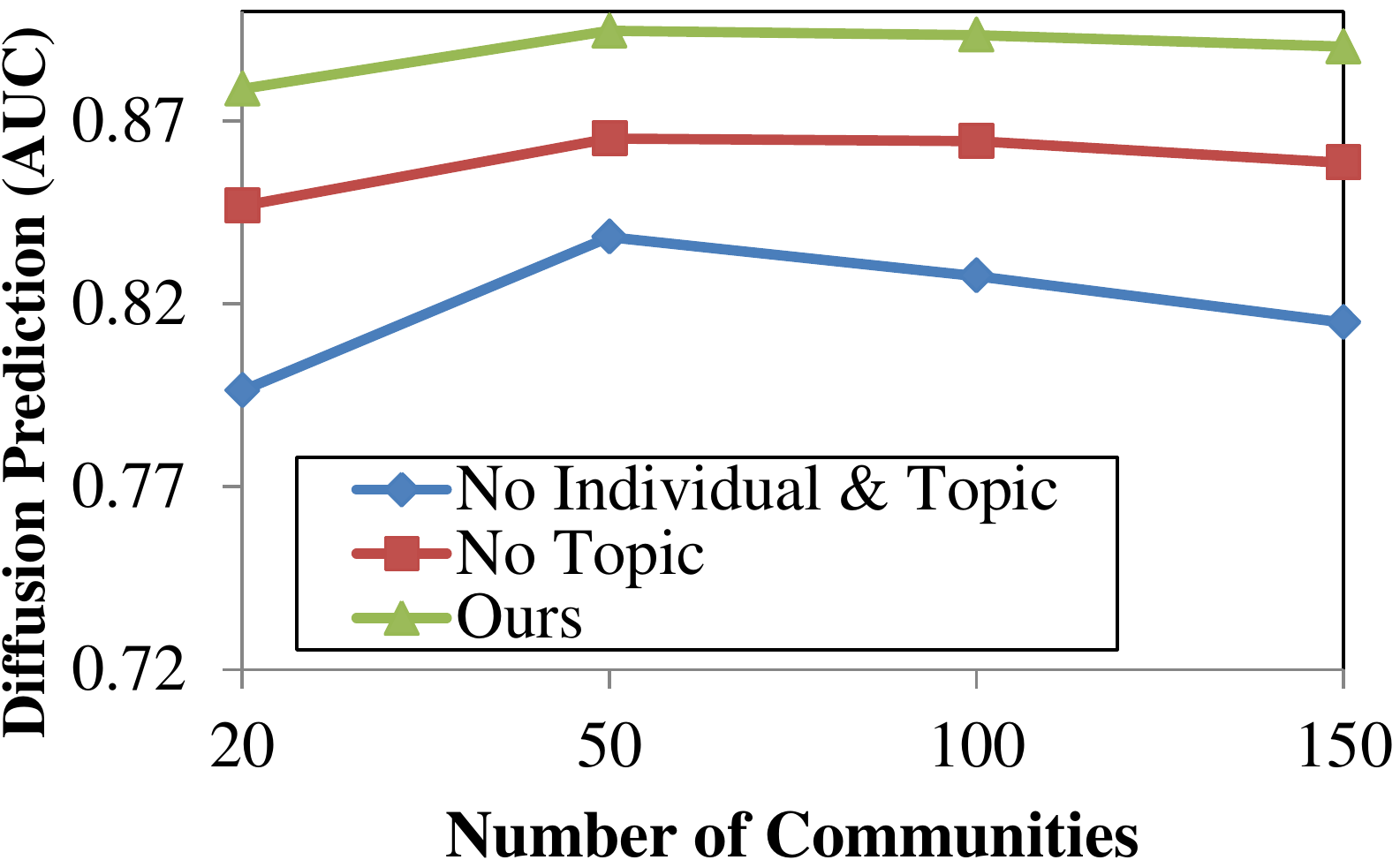}}
    \subfigure[Diffusion link pred. (DBLP)]{\label{fig:dblpcdm}
        \includegraphics[width=0.24\linewidth]{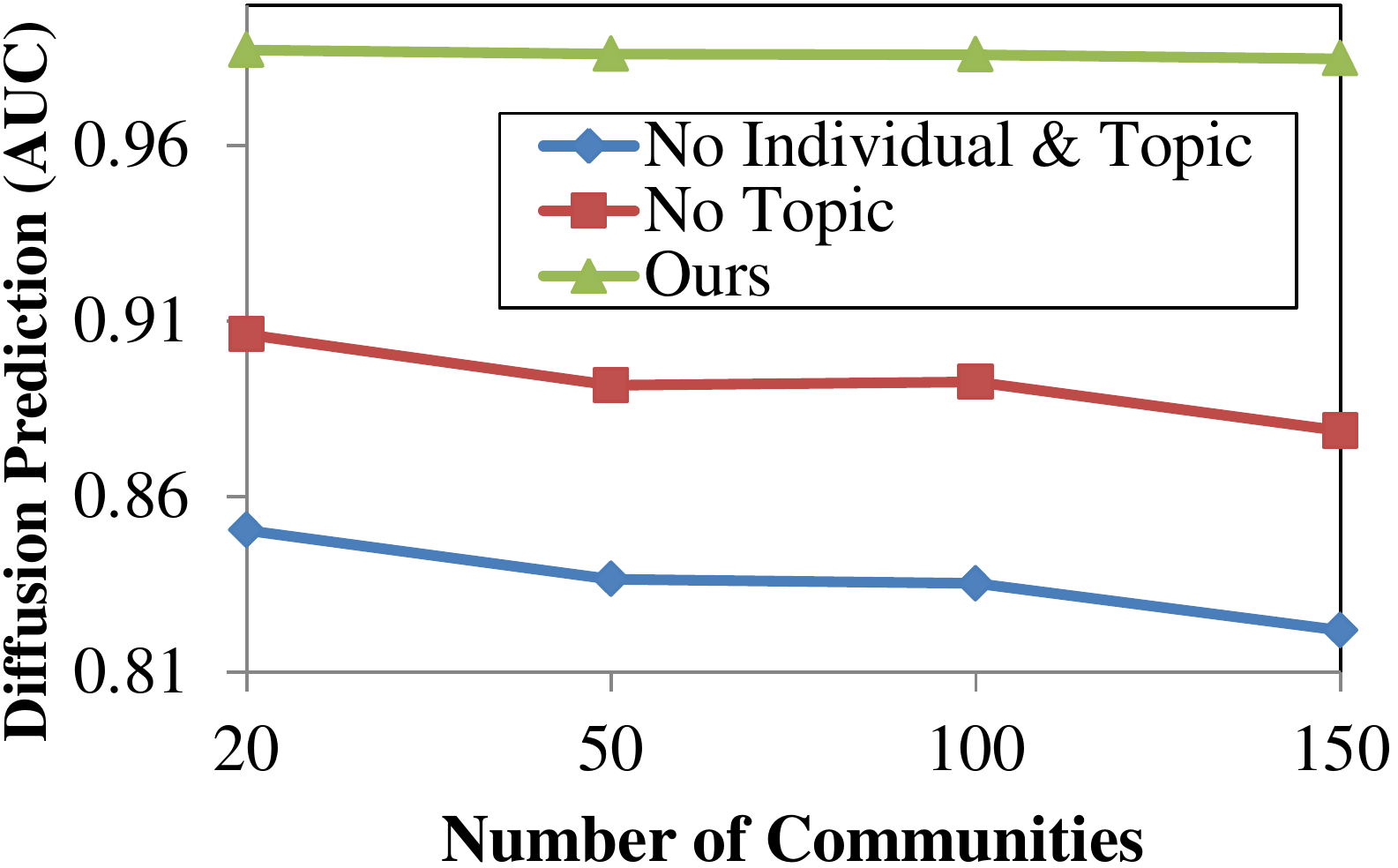}}
\end{tabular}
\vspace{-0.4cm}
\caption{Study of our model design.} \label{fig:ttwoinsights}
\end{figure*}

In addition to the above \emph{existing} baselines, we also design some more baselines to validate that we are better than a straightforward  community profiling approach of ``first detecting communities, then aggregating each community's user observations''. 
Specifically, we adopt the two state-of-the-art algorithms, CRM \cite{conf/kdd/HanT15} and COLD \cite{HuYCX15} to detect the communities, and further aggregate the user observations in each detected community as the profiles. 
After applying CRM and COLD, we get the community assignment probabilities for each user $u$ to each community $c$, which we denote as $\pi^*_{u,c}$'s. 
To get aggregated content profile, we first run LDA \cite{BleiNJ03} on all the users' documents with $|Z|$ topics, and for each user $u$'s $i$-th document $d_{ui}$, we get its $|Z|$-dimensional multinomial topic distribution as $\boldtheta^*_{d_{u,i}}$. Denote community $c$'s aggregated content profile as $\boldtheta^*_c$. We have 
\begin{equation} \label{eq.agg_content}
  \boldtheta^*_c \textstyle = \sum_{u=1}^{|U|} \pi^*_{u,c} \sum_{i = 1}^{|D_u|} \frac{\boldtheta^*_{d_{u,i}}}{|D_u|}.
\end{equation}
To get aggregated diffusion profile, we aggregate each diffusion link between $d_{u,i}$ and $d_{v,j}$ in $E$ w.r.t. their users' communities on a topic $z$. Denote aggregated diffusion profile $\eta^*_{c,c'z}$ as the probability of community $c$ diffusing community $c'$ on topic $z$. Then we have
\begin{equation} \label{eq.agg_diffusion}
  \eta^*_{c,c'z} \textstyle \propto \sum_{(i,j) \in E} \pi^*_{u,c} \pi^*_{v,c'}  \theta^*_{d_{u,i},z} \theta^*_{d_{v,j}, z}.
\end{equation}
In all, we obtain two more baselines, which implement the straightforward ``first detection, then aggregation'' profiling approach. 
\vspace{-2mm}
{\flushleft $\bullet$ \emph{CRM+Agg}.} 
It uses CRM \cite{conf/kdd/HanT15} to detect communities; then it uses Eq.~\ref{eq.agg_content} and Eq.~\ref{eq.agg_diffusion} for user aggregation to get the community content profiles and diffusion profiles, respectively. 
\vspace{-2mm}
{\flushleft $\bullet$ \emph{COLD+Agg}.} 
It uses COLD \cite{HuYCX15} to detect communities, then similarly uses Eq.~\ref{eq.agg_content} and Eq.~\ref{eq.agg_diffusion} to get content and diffusion profiles. 

We compare with both CRM+Agg and COLD+Agg in diffusion link prediction and community ranking. 

\vspace{0.05in}
\noindent \textbf{Evaluation}.
Since we jointly profile and detect communities, we will evaluate the quality of community detection and profiles.
\vspace{-2mm}
{\flushleft $\bullet$ \it{Detection quality}.} 
We consider two ways to evaluate the resulting communities:
1) how dense they are;
2) how well they can be used to explain the friendship link observations. 
For 1), we use \emph{conductance} \cite{Kloster:2014,LeskovecLM10} as the metric. 
As our community assignment is probabilistic, we follow \cite{HuYCX15} to let each user belong to her top five communities in conductance evaluation (and also later community ranking evaluation). The smaller conductance is, the better. 
For 2), we follow \cite{HuYCX15} to design a \emph{link prediction} task, where we use Eq.~\ref{equ:puv} to predict whether to observe a friendship link based on two users' communities. As there is no predefined threshold for link prediction, we use \emph{AUC} (Area Under the receiver operating characteristic Curve) \cite{HuYCX15,conf/kdd/HanT15} as the metric. 
Given a ranking of non-observed links, we calculate the AUC score as the probability of a randomly chosen true positive link being ranked higher than a randomly chosen true negative link. In the 10-fold cross validation, each time we use 10\% of the positive links and sample the same amount of negative links to calculate AUC. The higher AUC is, the better.
\vspace{-2mm}
{\flushleft $\bullet$ \it{Profile quality}.} 
Due to lack of ground truth, we generally evaluate the content and diffusion profiles' quality through the applications in Sect.~\ref{sec:applications}. 
For community-aware diffusion, as there is no predefined threshold for diffusion link prediction, we again use \emph{AUC} as the evaluation metric. 
For profile-driven community ranking, as the communities detected by different algorithms are different, for fair comparison, we evaluate the quality of each ranked community in terms of its users-- given a query $q$, we check how many users in the top $K$ ranked communities really retweet (or cite) about $q$. Then naturally we compute \emph{precision} and \emph{recall} for each community in the ranking list. Denote the users who mention $q$ in their retweets (or citation paper titles) as $U^*_q$. Denote the users belonging to any of the top $K$ communities as $U_K$. The precision of the top $K$ communities for query $q$ is $P(K,q) = |U^*_q \cap U_K| / |U_K|$, and the recall is $R(K,q) = |U^*_q \cap U_K| / |U^*_q|$. 
We define \emph{mean average precision} (MAP) over all the queries as $MAP@K = \sum_q (\sum_{i=1}^K P(i,q) / K) / |Q|$ and \emph{mean average recall} (MAR) as $MAR@K = \sum_q (\sum_{i=1}^K R(i,q) / K) / |Q|$.
Finally, we define the \emph{mean average F1} as $MAF@K = \frac{2 \times MAP@K \times MAR@K}{ MAP@K + MAR@K}$. 
The higher MAF is,  the better.
In addition, as the content profile is based on topics, we adopt one extra widely used metric (\emph{perplexity}) in topic modeling \cite{BleiNJ03} to evaluate its quality. Effectively, perplexity of a content profile measures how well it generates the user content observations, and we use the same definition of perplexity as in \cite{HuYCX15}. The lower perplexity is, the better.

\subsection{Model Design} \label{exp:threeinsights}

We want to evaluate how well we address each community profiling challenge as introduced in Sect.~\ref{sec:introduction}. 
To achieve this goal, we design some baselines based on the degenerated versions of CPD, for validating the advantages of our model design. We compare CPD with these baselines, and evaluate the quality of detected communities and profiles through three tasks: community detection, friendship link prediction and diffusion link prediction. 

\vspace{-2mm}
{\flushleft $\bullet$ \it{Modeling the inter-dependency with community detection}.} 
We design a baseline ``no joint modeling'', where we first detect communities only from the friendship links through a generative model by Eq.~\ref{equ:puv}, then we extract the profiles through a generative model as in CPD except having the communities fixed. 
As shown in Figures \ref{fig:twitterhjcon}--\ref{fig:dblphjdp}, ours is always better than ``no joint modeling''. 

\vspace{-2mm}
{\flushleft $\bullet$ \it{Addressing the heterogeneity of social observations}.} 
We design a baseline ``no heterogeneity'', where we adapt CPD to model friendship links and diffusion links in the same way by Eq. (\ref{equ:puv}), but keep the other parts of CPD modeling unchanged. As shown in Figures \ref{fig:twitterhjcon} - \ref{fig:dblphjdp}, ours is better than ``no heterogeneity'' on diffusion prediction, and comparable with it on community detection and friendship link prediction. This implies: 1) diffusion links and friendship links are different, and diffusion links require more sophisticated modeling than friendship links; 2) friendship links and diffusion links are correlated; diffusion links do not significantly change the community structure once the friendship links are given. 

\vspace{-2mm}
{\flushleft $\bullet$ \it{Accommodate the nonconformity of user behaviors}.}
We design two baselines: 
1) ``no individual \& topic'', where we exclude the individual factor and topic factor from Eq.~\ref{equ:pijt} in CPD;
2) ``no topic'', where we exclude only the topic factor from Eq.~\ref{equ:pijt} in CPD. 
As shown in figures \ref{fig:twittercdm} and \ref{fig:dblpcdm}, the individual factor is able to contribute 4.8\% and 6.8\% absolute AUC improvement on Twitter and DBLP respectively; the topic factor is able to contribute another 3.6\% and 10.5\% absolute AUC improvement on each data set. 

In all, we conclude that our model design well addresses the three challenges in community profiling.

\subsection{Comparison with Baselines} \label{subsec:cp}
We evaluate CPD and the baselines on various applications. 

\begin{figure}[t] 
\begin{tabular}[t]{l} \hspace{-0.4cm}
    \subfigure[Twitter]{\label{fig:twitterdp} 
        \includegraphics[width=0.49\linewidth]{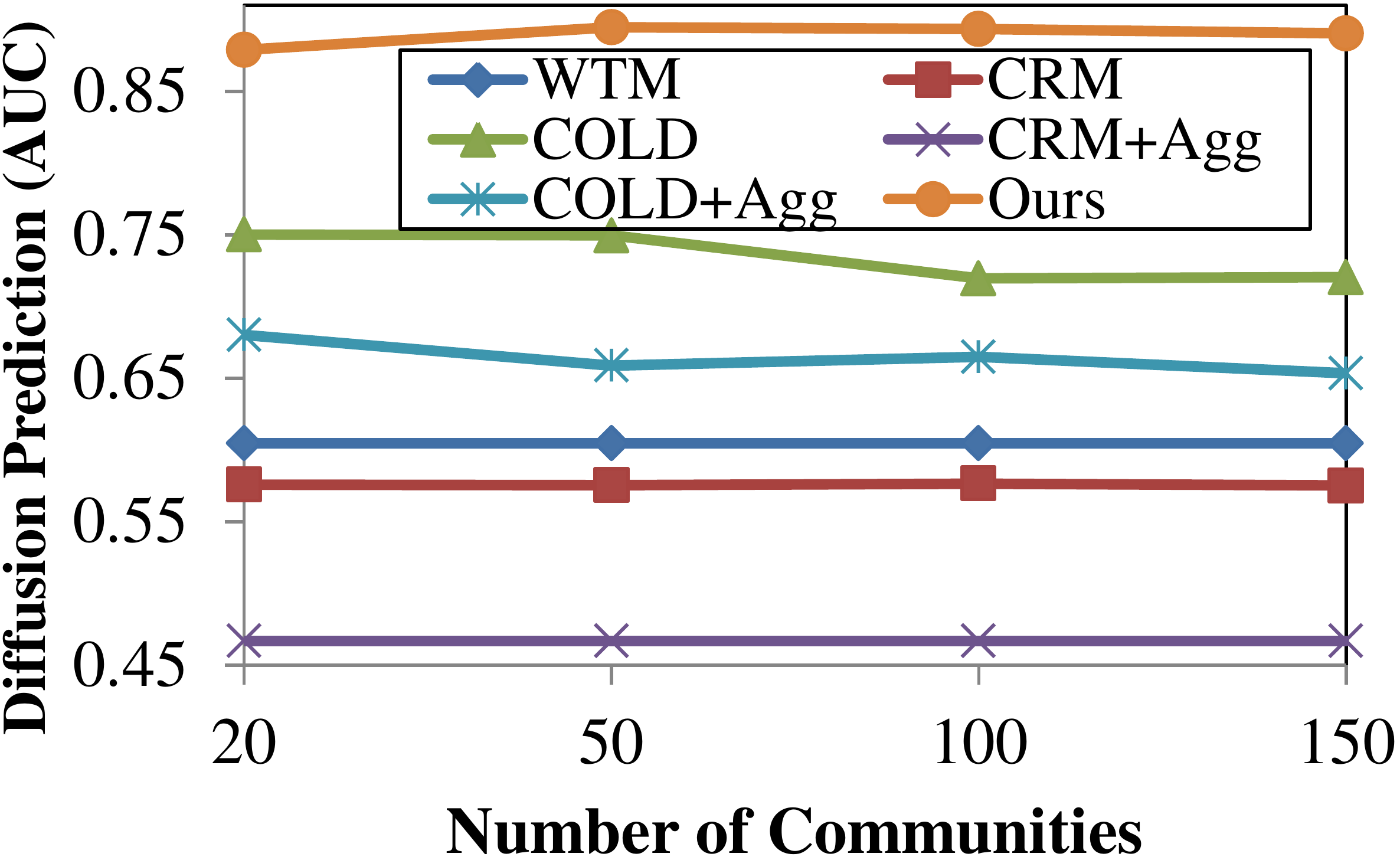}} \hspace{0.1cm}
    \subfigure[DBLP]{\label{fig:dblpdp}
        \includegraphics[width=0.49\linewidth]{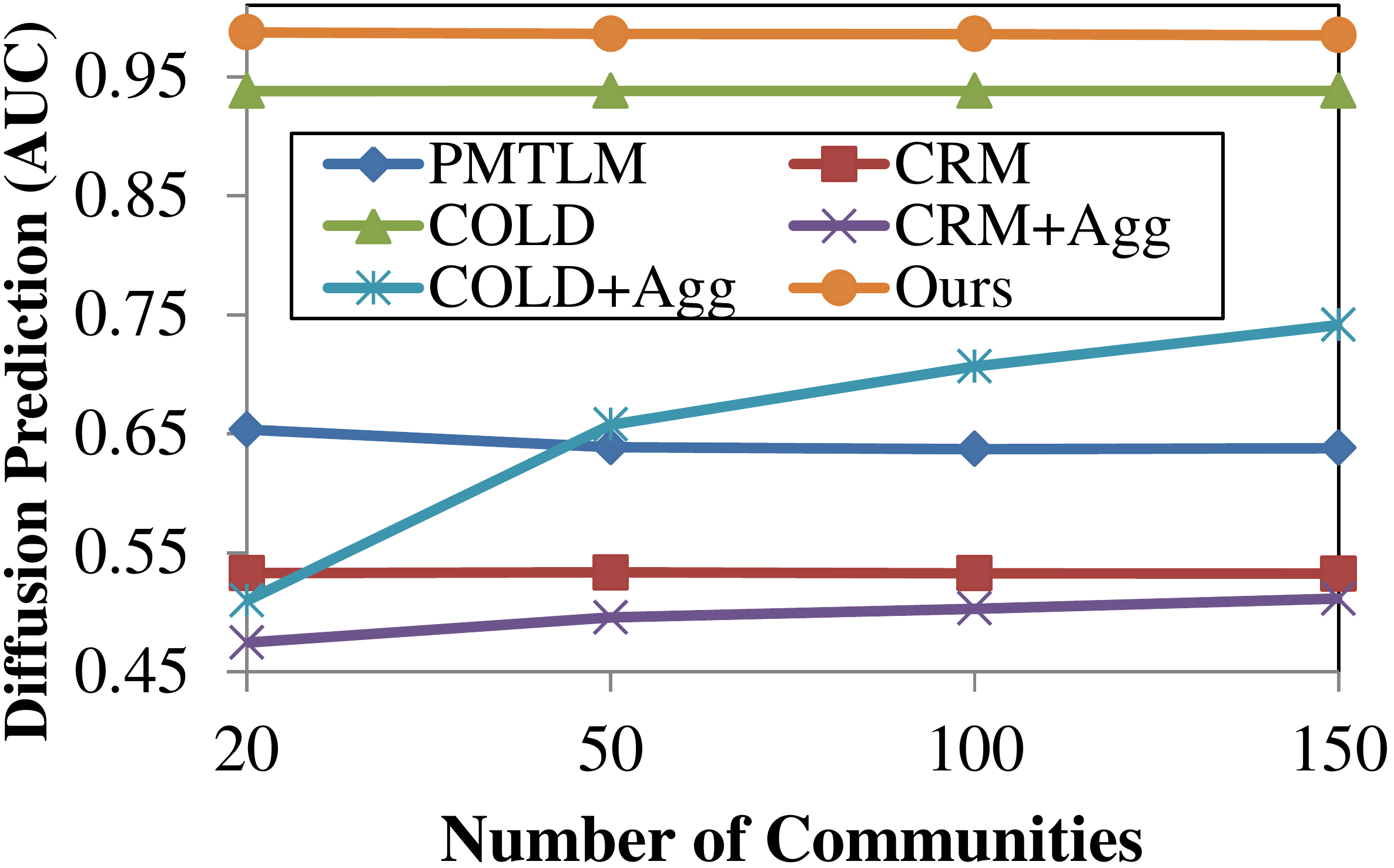}}
\end{tabular}
\vspace{-0.5cm}
\caption{Results of community-aware diffusion.} \label{fig:dpp}
\vspace{-0.2cm}
\end{figure}

\subsubsection{Community-aware Diffusion} \label{sec:cd}

\noindent\textbf{Quantitative analysis}. 
In figure \ref{fig:dpp}, we summarize the result comparison with the baselines introduced in Sect.~\ref{sec:setup}. 
PMLTM is not applicable to Twitter, since it is designed solely for citation network-- it predicts a citation based on the similarity between two documents, but in Twitter a tweet and its retweet are almost identical. 
As shown in Fig.~\ref{fig:dpp}, our model consistently outperforms all the baselines, thanks to: 1) our modeling various diffusion factors and heterogeneous user links, in contrast with the baselines in Table \ref{tab:comparedmethods}; 2) our joint detection and profiling, in contrast with the two ``first detection then aggregation'' baselines.  
When $|C|=100$, we achieve 24.2\%--91.6\% and 5.1\%--108.0\% relative AUC improvements than the baselines in Twitter and DBLP, respectively. The improvements are statistically significant over the 10-fold cross validation results, with student's $t$-test one-tailed $p$-value $p< 0.01$. 

\begin{figure}[t] 
\begin{tabular}[t]{c} \hspace{-6mm} 
    \subfigure[Individual factor]{\label{fig:user_factor} 
        \includegraphics[width=0.22\textwidth]{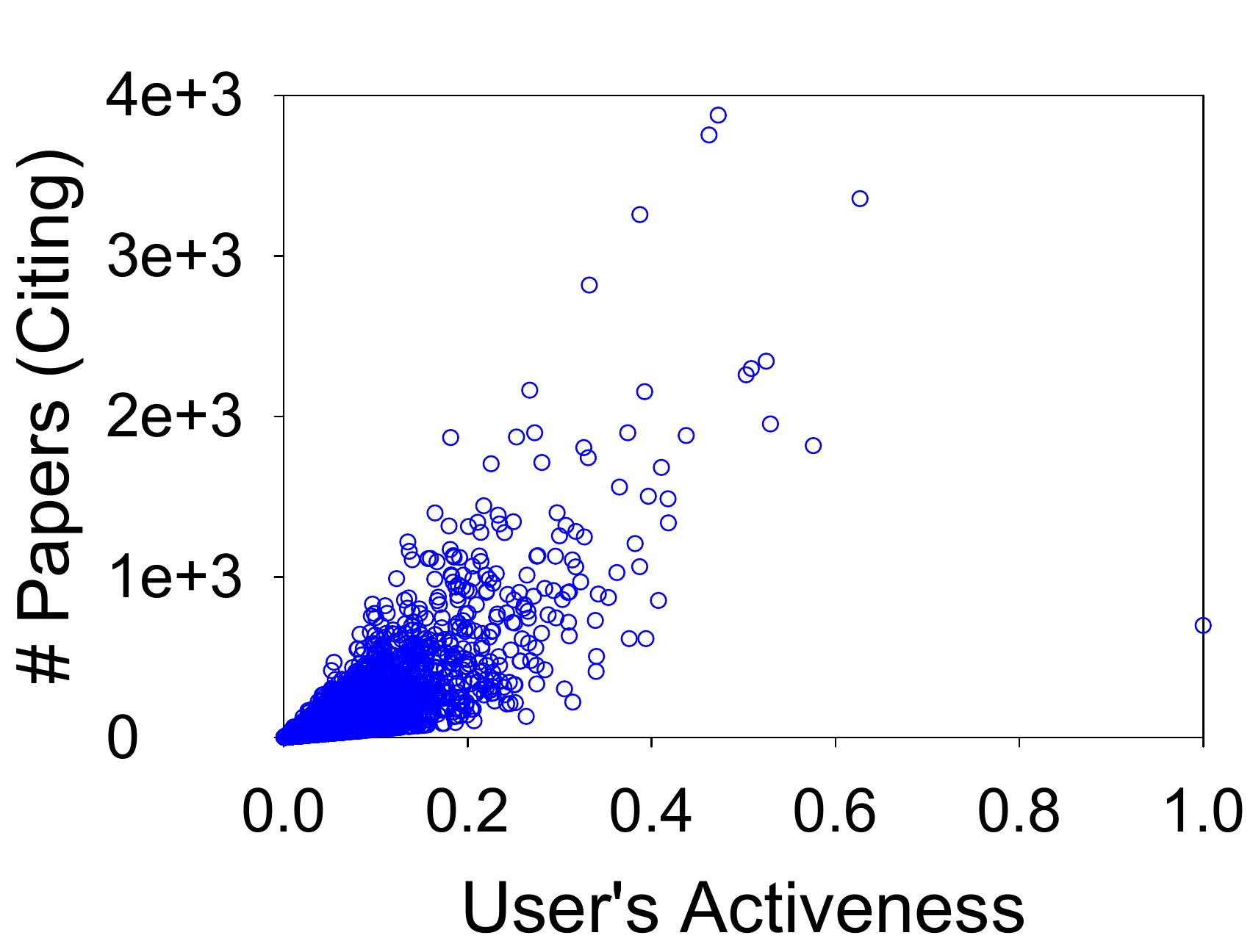}
        \includegraphics[width=0.22\textwidth]{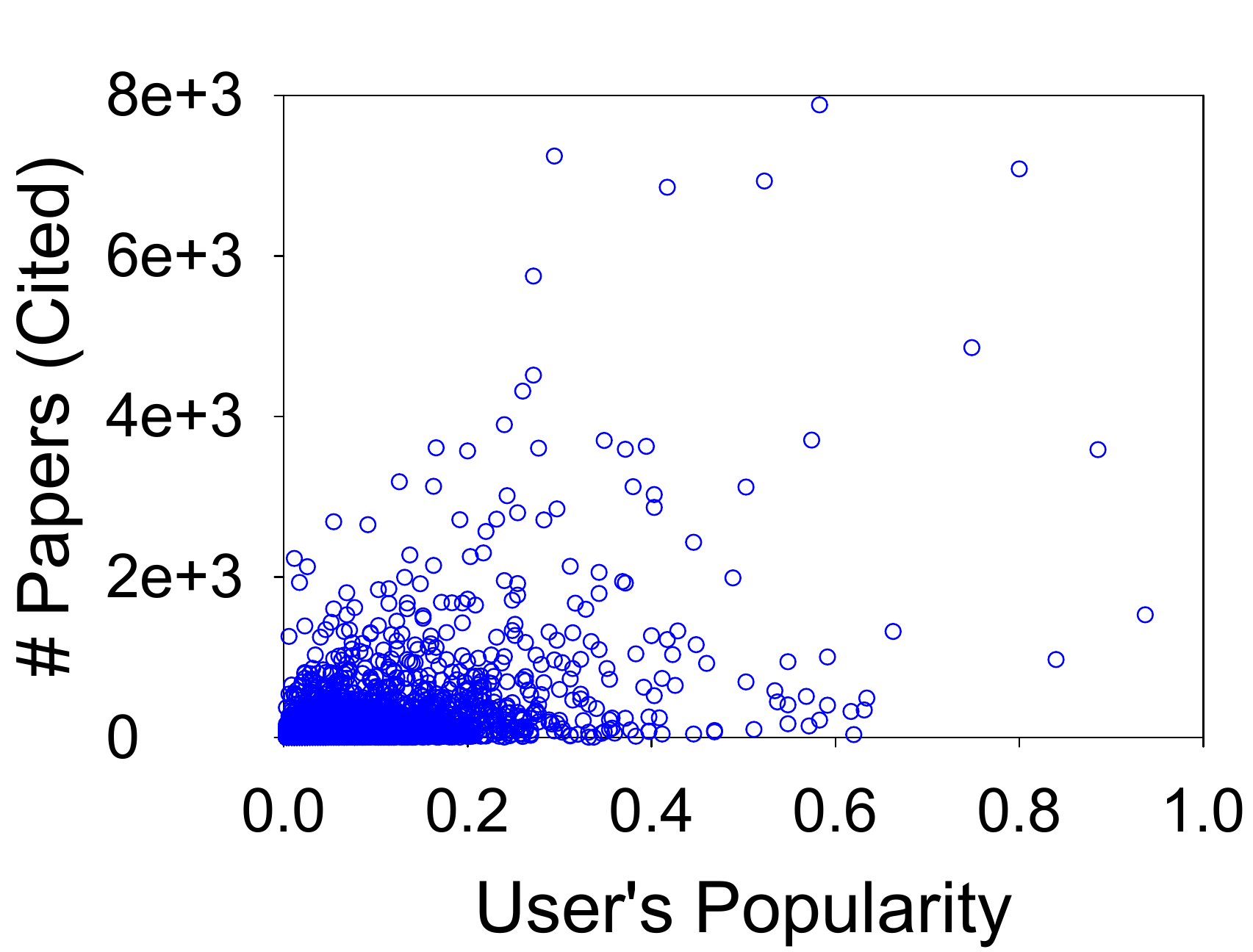}} \\ \vspace{-7.5mm} \\
     \subfigure[Topic factor]{\label{fig:topic_factor} \hspace{-3mm}
        \includegraphics[width=0.21\textwidth]{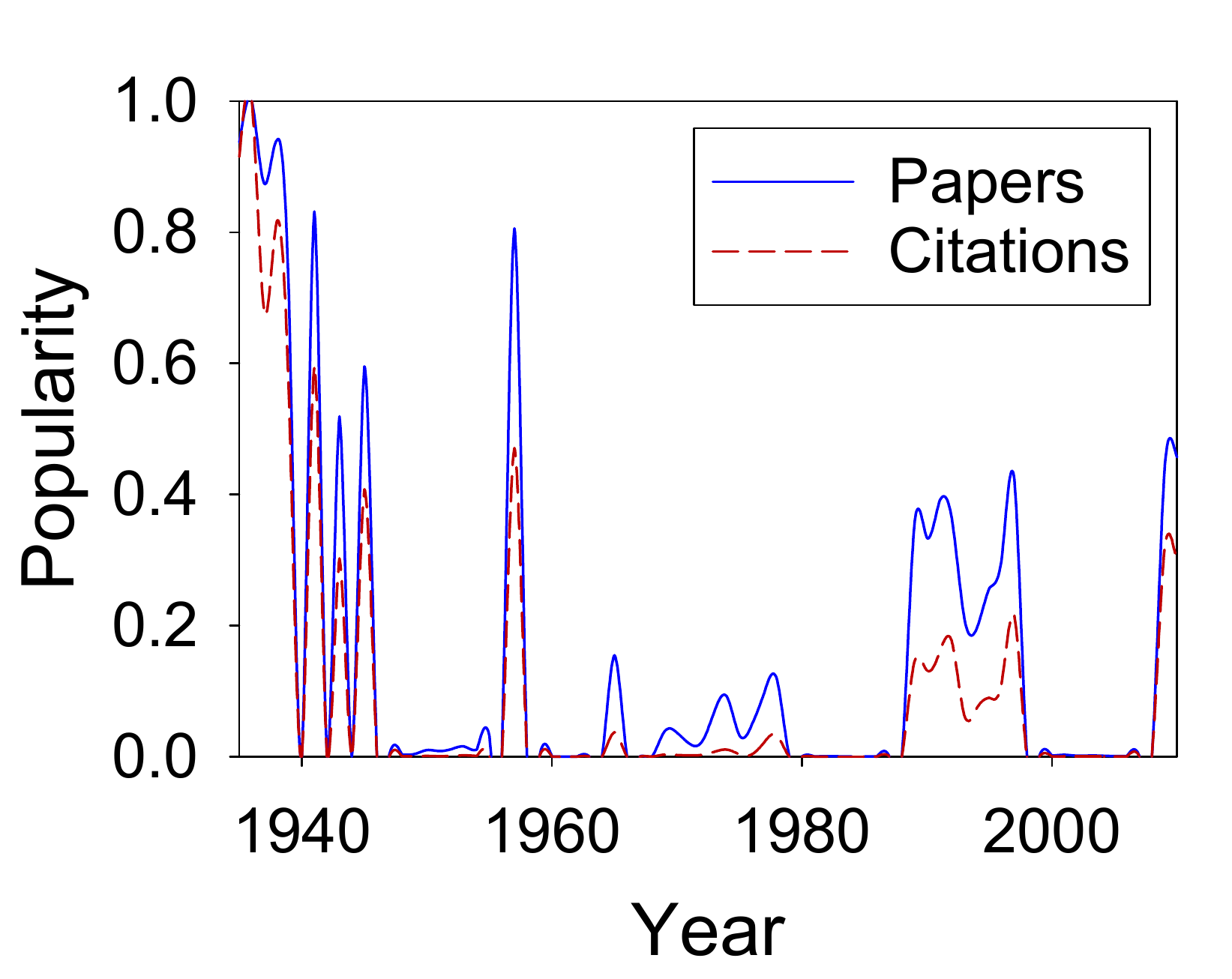}}  \hspace{1mm} 
     \subfigure[Community factor]{\label{fig:community_factor}
        \includegraphics[width=0.22\textwidth]{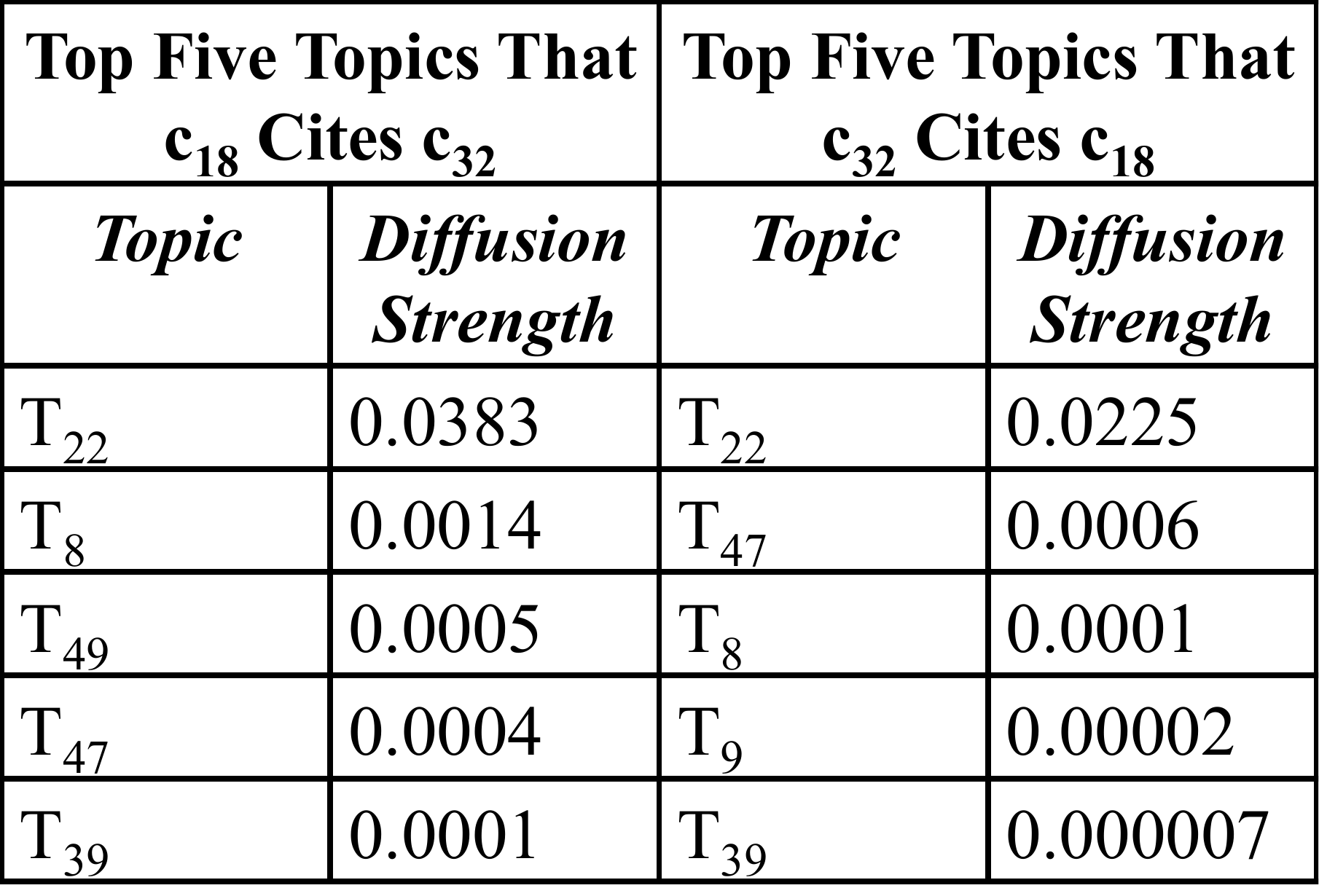}}    
\end{tabular}
\vspace{-2mm}
\caption{Community-aware diffusion case study.} \label{fig:indicasestudy}
\end{figure}

\begin{figure*}[t]
\small
\centering
\begin{tabular}[t]{c}  \hspace{-0.3cm}
    \subfigure[$|C|=50$ on Twitter]{\label{fig:crtwitter50}
        \includegraphics[width=0.24\textwidth]{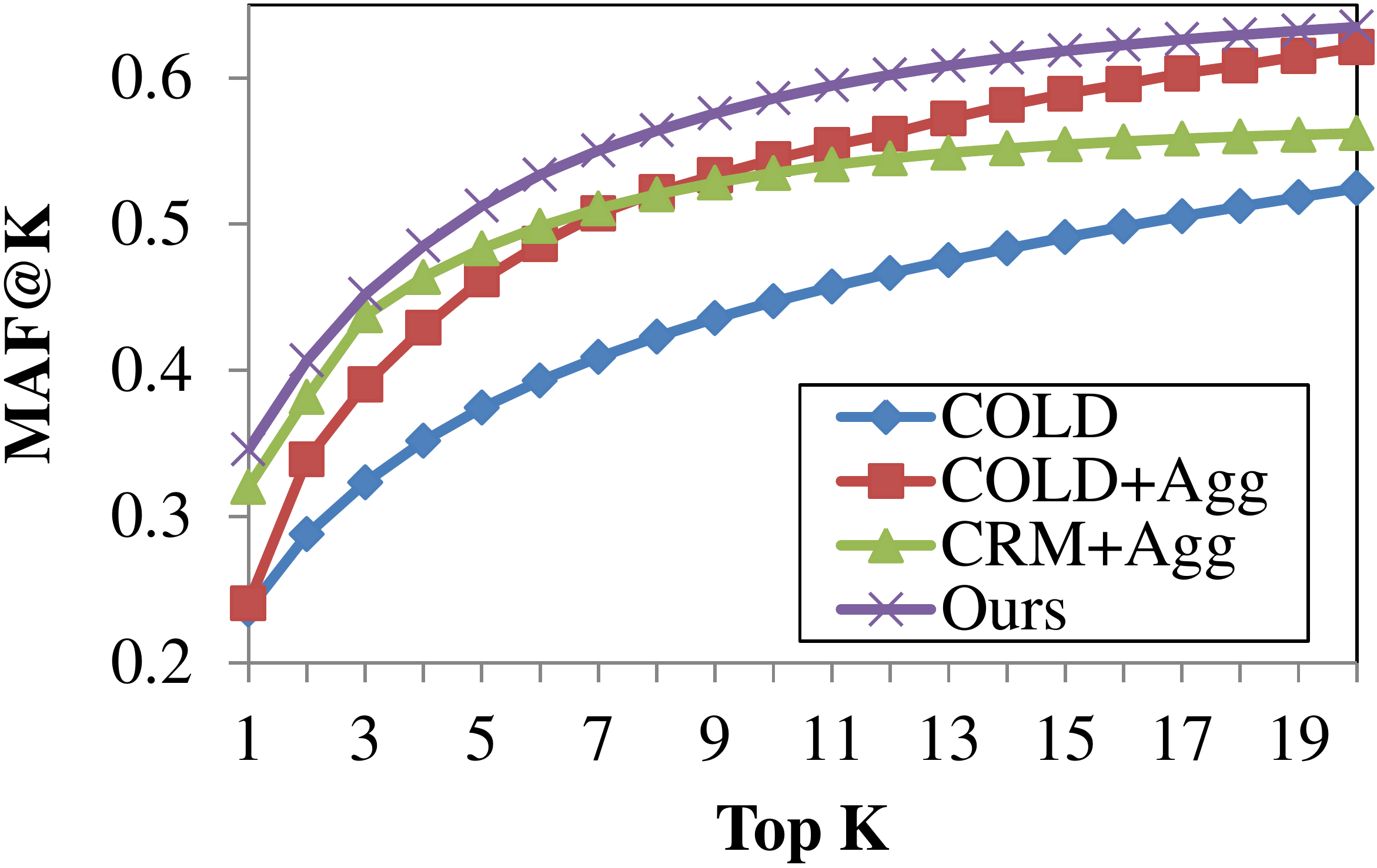}}
    \subfigure[$|C|=100$ on Twitter]{\label{fig:crtwitter100}
        \includegraphics[width=0.24\textwidth]{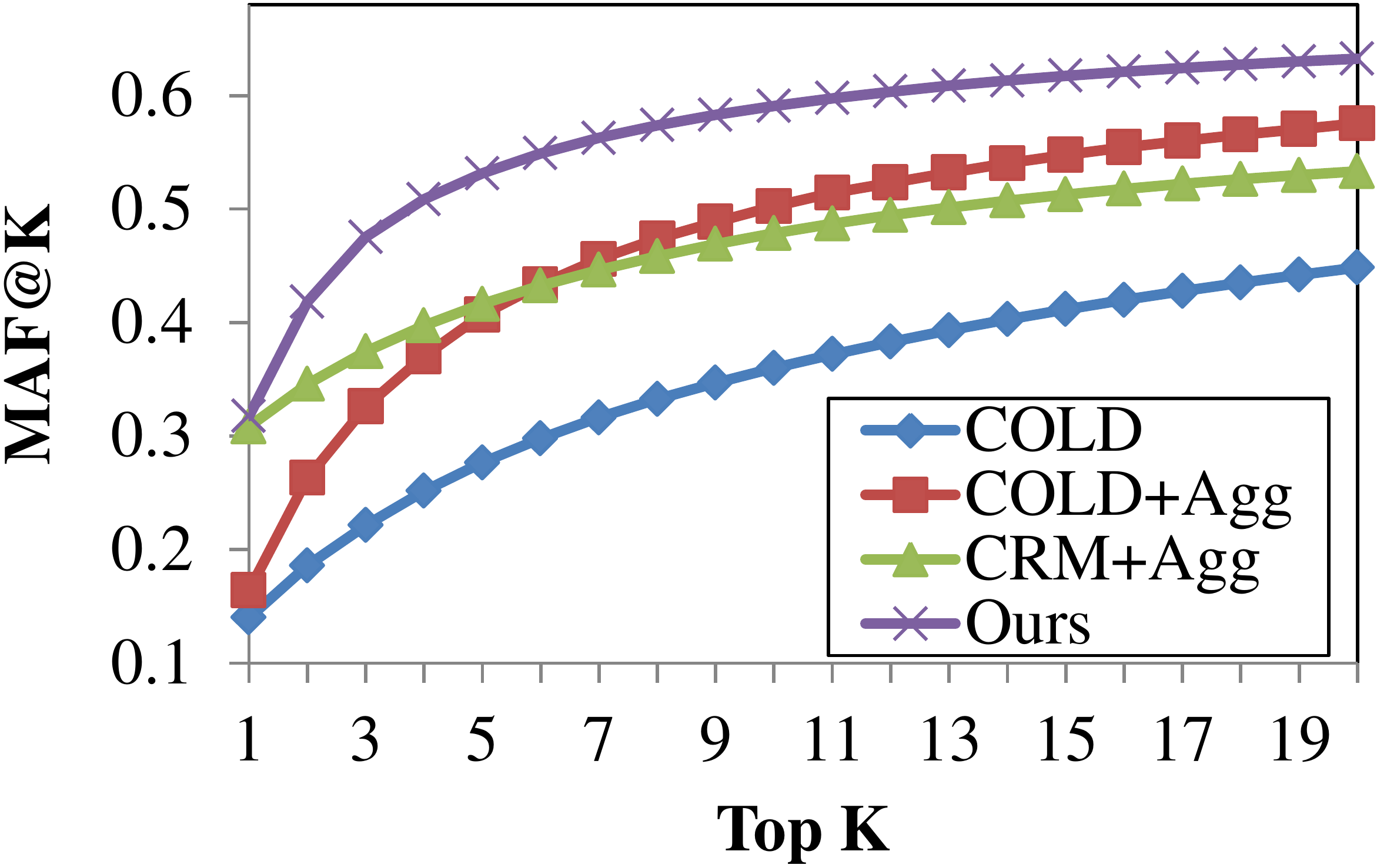}} 
    \subfigure[$|C|=50$ on DBLP]{\label{fig:crdblp50}
        \includegraphics[width=0.24\textwidth]{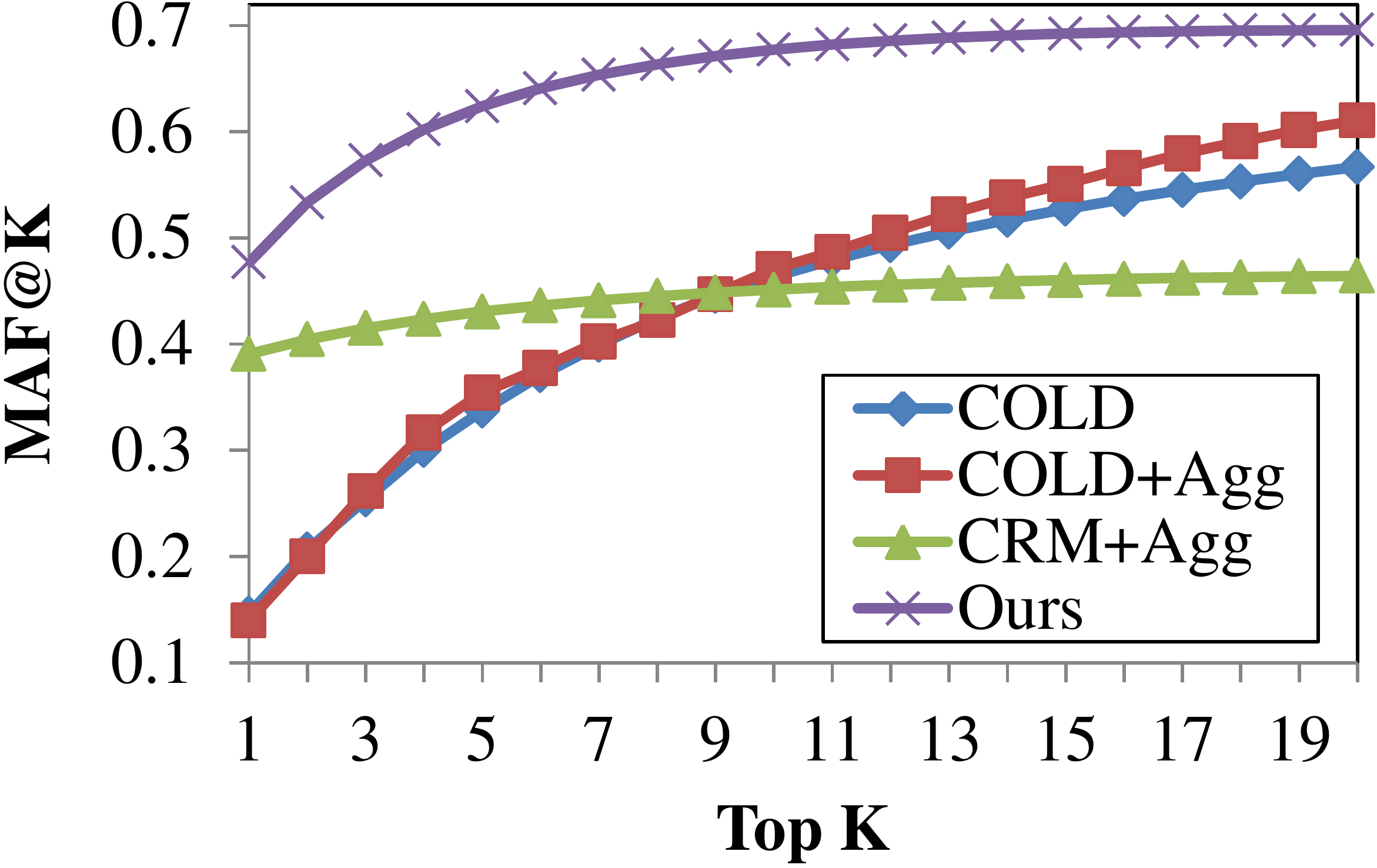}} 
    \subfigure[$|C|=100$ on DBLP]{\label{fig:crdblp100}
        \includegraphics[width=0.24\textwidth]{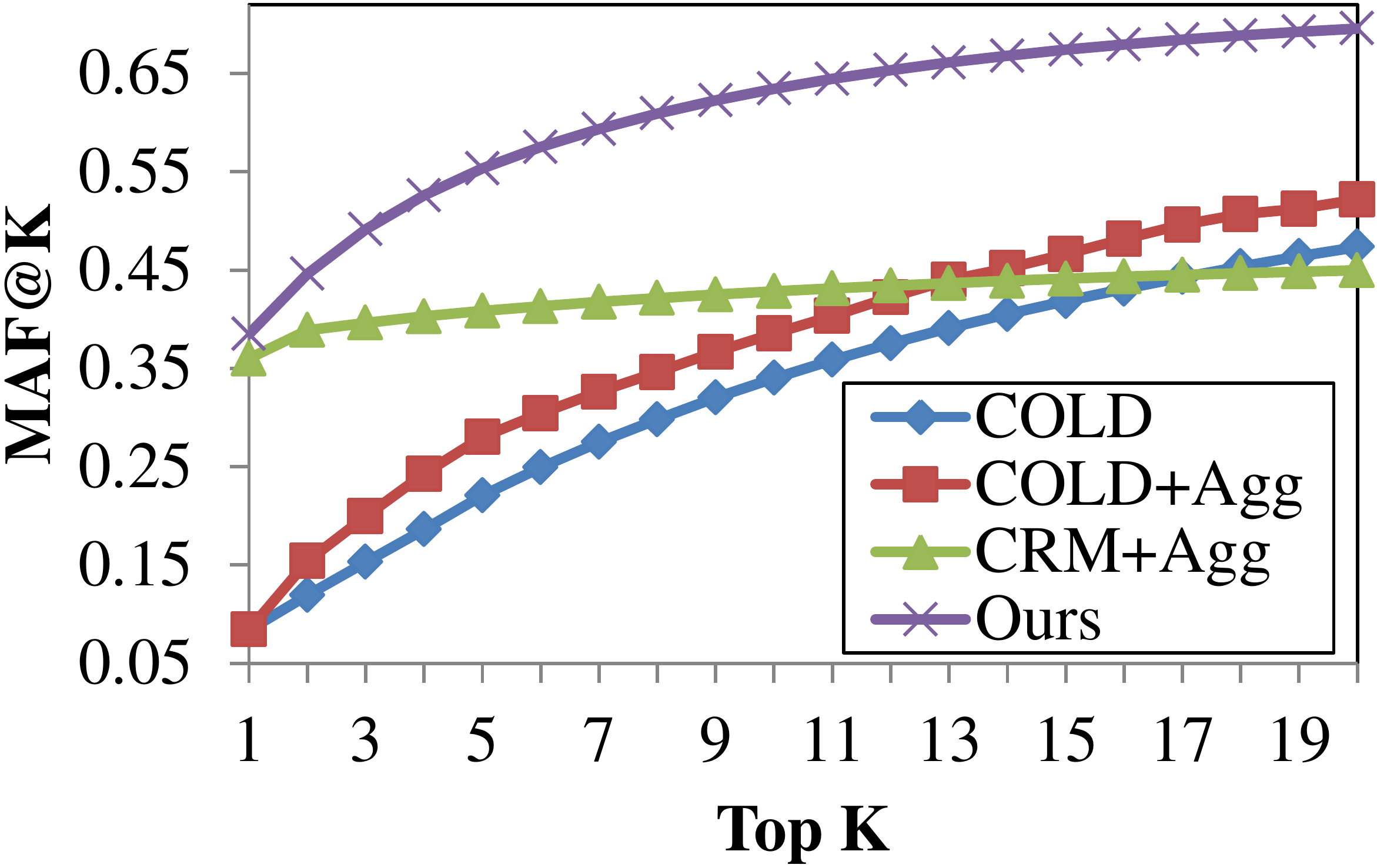}}        
\end{tabular}
\vspace{-0.4cm}
\caption{Results of profile-driven community ranking.} \label{fig:crp}
\end{figure*}

\vspace{0.05in}
\noindent\textbf{Case study}. We examine the three diffusion factors in Eq.~\ref{equ:pijt} with the DBLP data. 
Firstly, in Fig.~\ref{fig:user_factor} we plot the number of papers a user cites w.r.t. her \emph{activeness}, and the number of citations a user has w.r.t. her \emph{popularity}. User activeness and popularity are defined in Sect.~\ref{sec:modeldesign}. 
Generally, the more active a user is (i.e., publishing more papers), the more papers she cites; besides, the more popular a user is (i.e., a more established researcher), the more citations her papers get. This observation supports our design of modeling both user activeness and  popularity as the individual factors in diffusion.

In Figure \ref{fig:topic_factor}, we plot the number of papers and the number of citations w.r.t. a specific topic (e.g., ``parallel performance memory'') over the years. As we can see, there is a high correlation between the number of papers and that of citations over time-- if a topic is popular (i.e., it has many papers), then it is more likely to be cited (i.e., it appears in many citations). This observation supports our design of modeling the topic factor in Sect. \ref{sec:modeldesign}.

Finally, in Fig.~\ref{fig:community_factor} we list the diffusion between two example communities : $c_{18}$ and $c_{32}$, which are the top 2 communities ranked for query ``router'' in profile-drive community ranking (Sect.~\ref{sec:cr}). 
As we can see, $c_{18}$ and $c_{32}$ tend to cite from each other on topic $T_{22}$ (i.e., ``network'' as shown in Table \ref{tab:crworddistribution}). Besides, $c_{18}$ tends to cite $c_{32}$ on $T_{8}$ (i.e., ``security''), whereas $c_{32}$ tends to cite $c_{18}$ on $T_{47}$ (i.e., ``service''). This observation means: each community has a preference to diffuse other communities on certain topics. Thus it is necessary to model the community factor in diffusion.

\begin{table}[t]
	\centering  
	\addtolength{\tabcolsep}{-3pt}
	\begin{small}
		\begin{tabular}{|l| l|}
			\hline
			{\bfseries Topic}& {\bfseries  Word Distribution (listed by ``word:probability'')}         \\ \hline
			$T_{22}$ & network:0.059, wireless:0.050, sensor:0.046, routing:0.038 \\
			$T_{49}$ &  network:0.042, performance:0.037, traffic:0.031, routing:0.028 \\ 
			$T_{47}$ &  service:0.056, web:0.028, mobile:0.025, management:0.024 \\
			$T_{8}$ & security:0.031, key:0.028, authentication:0.027, protocol:0.020 \\
			$T_{9}$ &  code:0.061, algorithm:0.032, function:0.028, linear:0.027\\
			$T_{0}$ & design:0.049, circuit:0.034, power:0.027, cmos:0.017 \\			
			$T_{44}$ & parallel:0.053, performance:0.036, memory:0.03, architecture:0.02  \\
			$T_{46}$ & analysis:0.061, reliability:0.029, optical:0.024, design:0.021 \\ 
			\hline
		\end{tabular}
	\end{small} 
	\vspace{-0.3cm}
	\caption{Top four words in each topic.} 
	\label{tab:crworddistribution}  
\end{table} 

\begin{table}[t]
	\centering  
	\addtolength{\tabcolsep}{-0.3pt}
	\begin{small}
		\begin{tabular}{|c|c|c|c|c|}
			\hline
			{\bfseries K}& {\bfseries  AP@K} & {\bfseries AR@K} & {\bfseries AF@K} & {\bfseries Topic Distribution}        \\ \hline
			1 & 0.919 & 0.327 & 0.483 & $T_{22}$:0.976, $T_{49}$:0.013, $T_{47}$:0.006\\
			2 & 0.900 & 0.424 & 0.576 & $T_{8}$:0.988, $T_{22}$:0.004, $T_{9}$:0.003\\
			3 & 0.891 & 0.528& 0.663 & $T_{0}$:0.977, $T_{44}$:0.008, $T_{46}$:0.005\\
			\hline
		\end{tabular}
	\end{small} 
	\vspace{-0.3cm}
	\caption{Top three communities ranked for query ``router".} 
	\vspace{-0.2cm}
	\label{tab:crcasestudy}  
\end{table} 

\subsubsection{Profile-driven Community Ranking} \label{sec:cr} 
For community ranking, we follow several guidelines to choose queries: 
1) it should be easy to assess whether a retweet or a citation contains a query, thus we choose single terms (i.e., either hashtags or words) as queries;
2) a query has to be meaningful-- since words are noisy, we choose hashtags as queries in Twitter; DBLP has no hashtag, thus we choose words as queries, but we remove the top 1,000 frequent words; 
3) a query has to appear with sufficient frequency in retweets or citations, thus we choose hashtags in Twitter and words in DBLP with frequency both larger than 100. 
In the end, we have 5,680 queries in Twitter and 27,479 queries in DBLP.

Given each query $q$, we rank the detected communities by Eq. (\ref{eq.community_ranking}), and then return the top $K$ (for $K = 1, ..., 20$).

\vspace{0.05in}
\noindent\textbf{Quantitative analysis}. Fig.~\ref{fig:crp} compares our model with the baselines that support community-level content and diffusion modeling, including COLD, COLD+Agg and CRM+Agg. 
As we can see, our model consistently outperforms all the baselines; when $|C|=100$ and $K = 5$, we achieve 27.6\%--92.0\% and 35.4\%--150.8\% relatively MAF improvements than the baselines in Twitter and DBLP, respectively. 
All these improvements are statistically significant over the 10-fold cross validation results, with student's $t$-test one-tailed $p$-value $p< 0.01$. 
Note that our model is better than COLD+Agg and CRM+Agg, again showing the advantage of joint detection and profiling. 
Besides, we observe that our model's MAF@K starts to converge earlier than the baselines. This means we are able to find more relevant users in the top $K$ communities. 

We also tested community ranking with different subsets of queries. We divided the queries according to their occurrence frequency in the corpus. We equally splitted the range from the minimal frequency and the maximal frequency into five intervals. For each interval, we tested community ranking with the subset of queries, whose frequency falls within that interval. We observed similar trends that our model consistently outperforms the baselines. We also observed that the absolute MAF@K values are not sensitive to different query subsets.

\vspace{0.05in}
\noindent\textbf{Case study}. We further examine the communities ranked by our model for a specific query. 
Table \ref{tab:crcasestudy} lists the top three communities that are most likely to cite papers about ``router". AP@K is the average precision of query ``router'' for the top $K$ communities; similarly, AR@K is the average recall and AF@K is the average F1. 
AF@K increases as $K$ increases, which is consistent with the trend observed in Fig.~\ref{fig:crp}. 
Besides, according to Table \ref{tab:crworddistribution}, the top three communities to cite ``router" are: ``network wireless sensor", ``security key authentication" and  ``circuits design", all of which are reasonablly the networking communities. 

\begin{figure*}[t]
\hspace{-8mm}
\begin{minipage}{0.67\textwidth}
\small
\centering
    \subfigure[Diffusion with topic aggregation]{\label{fig:cvall}
        \includegraphics[width=0.32\textwidth,height=21mm]{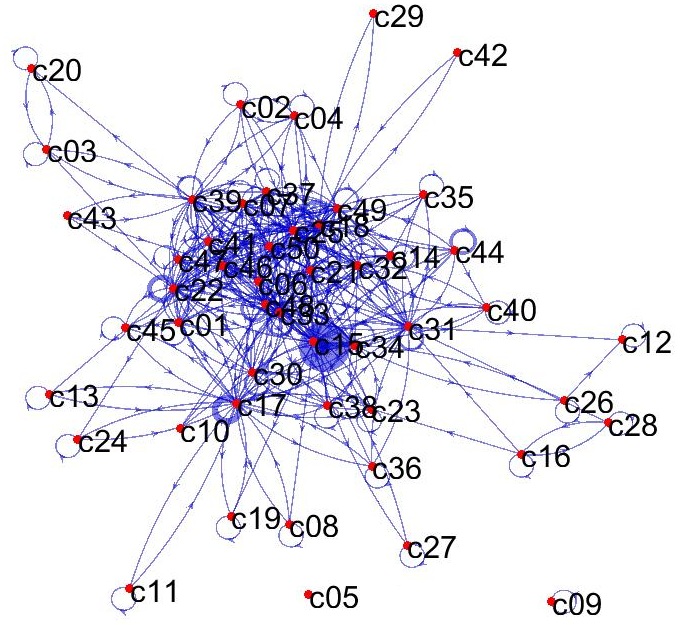}} \hspace{1mm}
    \subfigure[Diffusion on a general topic]{\label{fig:cv1}
        \includegraphics[width=0.26\textwidth,height=21mm]{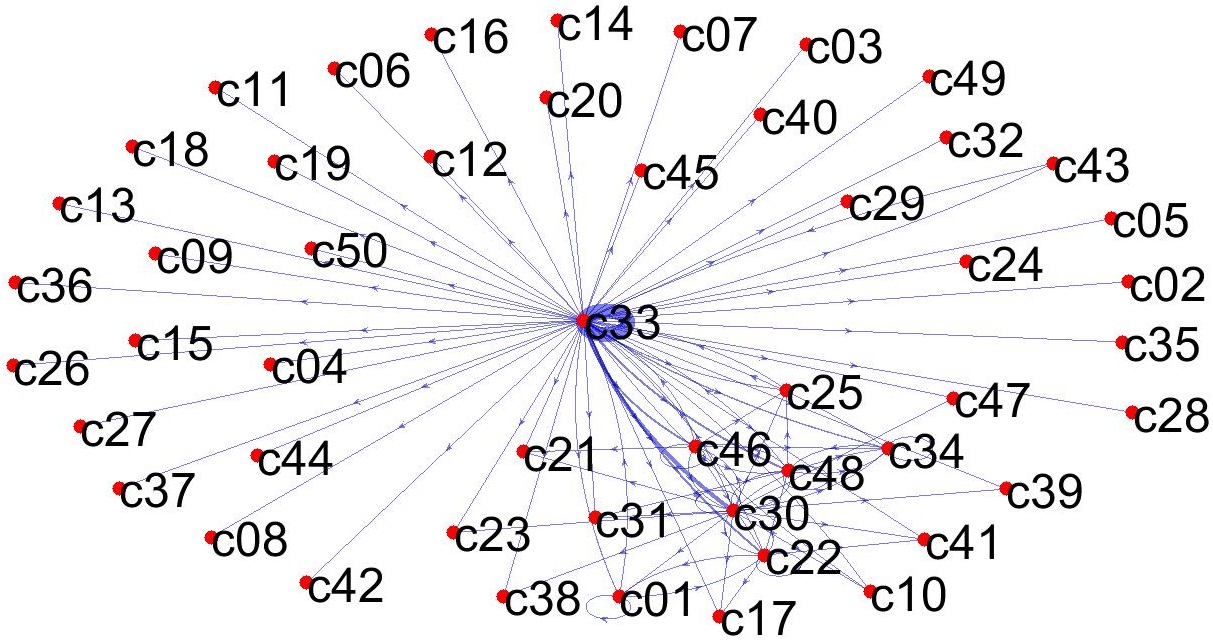}} \hspace{1mm}
    \subfigure[Diffusion on a specialized topic]{\label{fig:cv2}
        \includegraphics[width=0.24\textwidth,height=21mm]{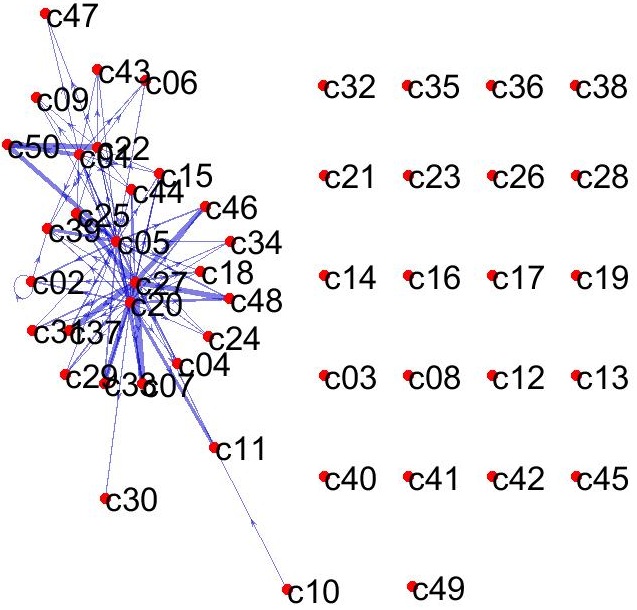}} 
\vspace{-4mm}
\caption{Results of profile-drive community visualization.} \label{fig:cv}
\end{minipage}
\hspace{-4mm}
\begin{minipage}{0.4\linewidth}
\centering
\addtolength{\tabcolsep}{-4pt}
	\begin{small}
		\begin{tabular}{|c|c|c|c|c|c|} 
			\hline
			\multirow{2}{*}{\bfseries Dataset}&\multirow{2}{*}{\bfseries Algorithms} &\multicolumn{4}{c|}{\bfseries Number of Communities} \\ \cline{3-6}
			&& {\bfseries  20} & {\bfseries 50} & {\bfseries 100} & {\bfseries 150}        \\ \hline
			\multirow{3}{*}{Twitter}&COLD+Agg & 825943.4 & 694741.7 & 515997.0 & 427181.1\\
			&CRM+Agg & 826737.5 & 695400.6 & 516730.9 & 427761.2\\
			&Ours & 5117.0 & 3992.1& 3801.4 & 3849.7\\ \hline
			\multirow{3}{*}{DBLP}&COLD+Agg & 61348.6 & 47179.0 & 39983.4 & 36922.7\\
			&CRM+Agg & 61901.9 & 47480.5 & 41289.1 & 37018.3\\ 
			&Ours & 1153.0 & 982.0 & 875.0 & 885.0\\ 
			\hline
		\end{tabular}
	\end{small} 
	\vspace{-2mm}
	\caption{Perplexity Comparison.} \label{tab:cppp}
\end{minipage}
\end{figure*}

\begin{figure*}
\small
\centering
\begin{tabular}[t]{c} 
    \subfigure[Community detect. (Twitter)]{\label{fig:twittercon} \hspace{-2mm}
        \includegraphics[width=0.24\textwidth]{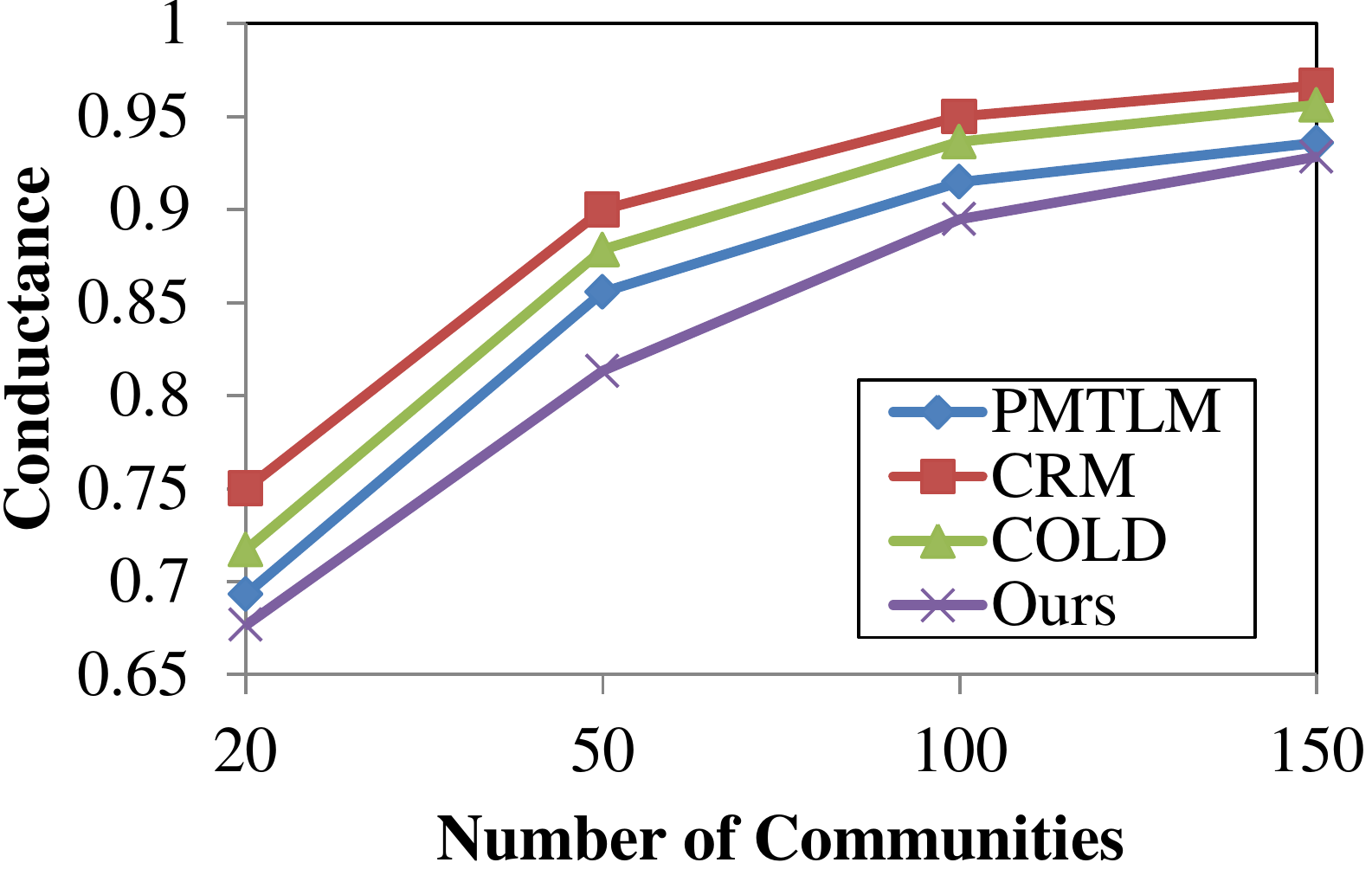}}
    \subfigure[Friendship link pred. (Twitter)]{\label{fig:twitterlp}
        \includegraphics[width=0.24\textwidth]{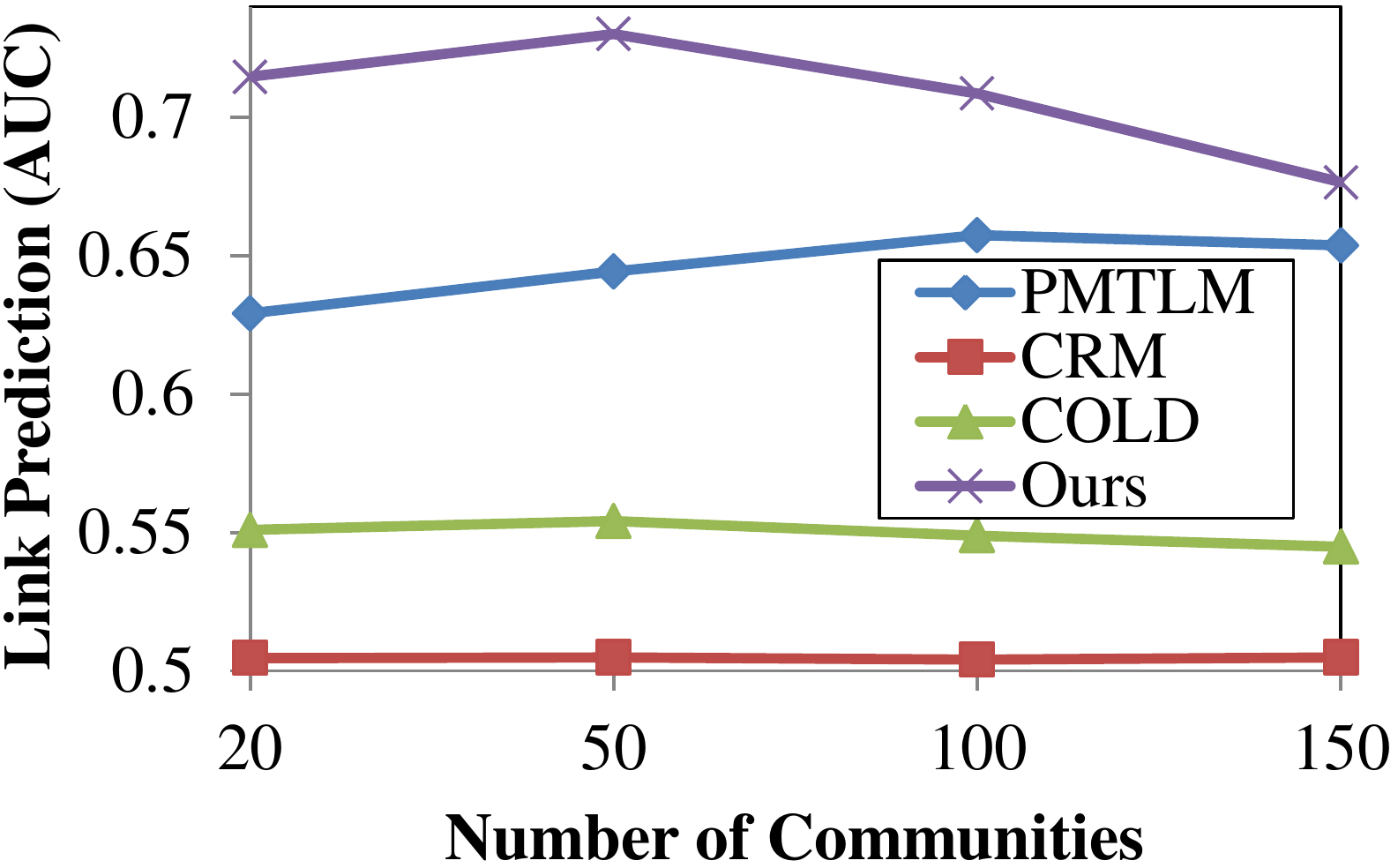}}
    \subfigure[Community detect. (DBLP)]{\label{fig:dblpcon}
        \includegraphics[width=0.24\textwidth]{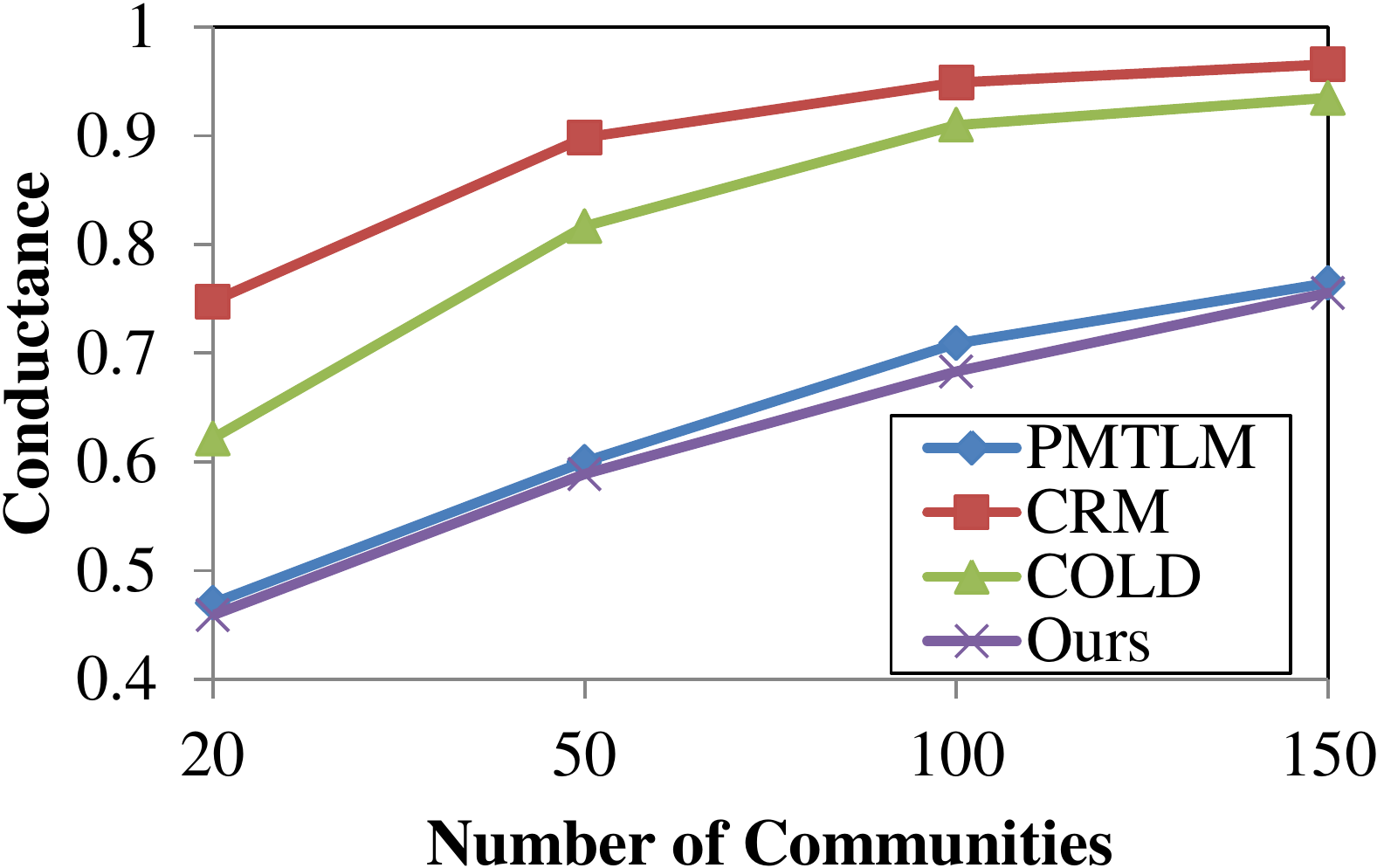}}
    \subfigure[Friendship link pred. (DBLP)]{\label{fig:dblplp}
        \includegraphics[width=0.24\textwidth]{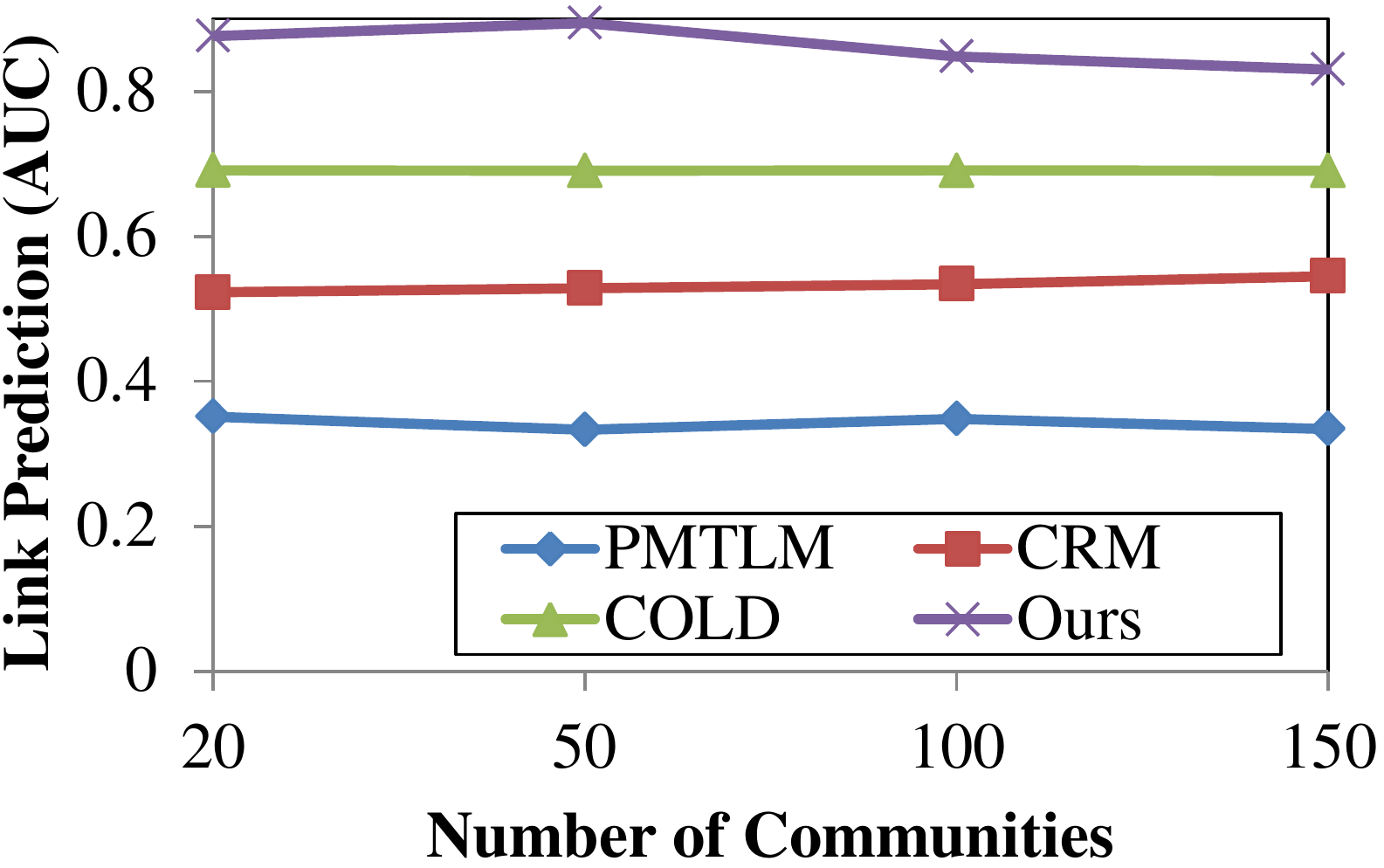}}        
\end{tabular}
\vspace{-5mm}
\caption{Results of community detection.} \label{fig:cdp}
\end{figure*}

\subsubsection{Profile-driven Community Visualization}
In Fig.~\ref{fig:cv}, we visualize the DBLP community diffusion under aggregation of all topics, a general topic and a specialized topic, respectively. In total, we detect 50 communities and denote them as $c_{01}$--$c_{50}$. For each directed edge between two communities $c$ and $c'$, the width indicates the diffusion strength. In Fig.~\ref{fig:cvall}, the strength is an aggregated value $\sum_z \eta_{cc'z}$ over all the topics; in Figures \ref{fig:cv1} and \ref{fig:cv2} the strength is $\eta_{c,c'z}$ for a specific topic $z$. We skip the edges whose strengths are below average for simpler visualization. 

We can make several interesting observations from Fig.~\ref{fig:cv}.
Firstly, in Fig.~\ref{fig:cvall} we find that, under topic aggregation, the communities often diffuse a lot within themselves. This coincides with our definition of ``community'' that the group of users who share similar diffusion behavior-- in this case, the same community users often diffuse information to each other. 
Secondly, in Fig.~\ref{fig:cvall} we also find that, some communities are more ``open'' than the others. E.g., $c_{48}$ (``data database search") and $c_{33}$ (``web information analysis") are more open research communities, which diffuse information with most of the other communities. In contrast, $c_{09}$ (``neural control system") appears as a more closed research community, which hardly diffuses information with other communities. Such a visualization enables us to assess the openness of a research community. 
Finally, we find that, the diffusion behaviors vary w.r.t. different kinds of topics. E.g., Fig.~\ref{fig:cv1} shows the diffusion on a very general topic (``web, information, search, semantic''), which can be discussed and diffused by many research communities. In contrast, Fig.~\ref{fig:cv2} shows the diffusion on a very specialized topic (``transmission, gbs, trail, video''), which is of interest to only a few communities such as $c_{25}$ (``distributed performance computing") and $c_{27}$ (``reliability device design"). This visualization reveals the topic generality and is helpful to researchers in choosing research topics.

\subsubsection{Quality of Community and Content Profile} 
In addition to the three applications, we also conduct experiments to evaluate the quality of communities and content profiles. In Fig.~\ref{fig:cdp}, we show our model consistently outperforms the baselines in terms of community quality. As COLD+Agg and CRM+Agg use the detection of COLD and CRM respectively, we do not include them in comparison again. When $|C|=100$, we achieve: 1) 2.2\%--5.8\% (Twitter) and 3.5\%--27.8\% (DBLP) relative conductance improvements; 2) 7.8\%--40.6\% (Twitter) and 22.8\%--143.5\% (DBLP) relative AUC improvements. All the improvements are significant with $p$-values $p< 0.01$. 
In general, we are better than COLD and PMLTM, as they do not model the friendship links in community detection; we are better than CRM, as it does not enforce dense friendship links in a community. 

In Fig.~\ref{tab:cppp}, we compare with COLD+Agg and CRM+Agg in terms of the quality of content profiles. 
As we can see, our model achieves the lowest perplexity, meaning that our content profiles can best explain the user content observations. This supports our argument of joint modeling, as motivated in Eq.~\ref{eq.content_profile}. 

\subsection{Scalability} \label{sec.scalability} 
\begin{figure}
\small
\begin{tabular}[t]{c} \hspace{-0.4cm}
   \subfigure[Time vs. data size]{ \label{fig:scalability}
        \includegraphics[width=0.48\linewidth]{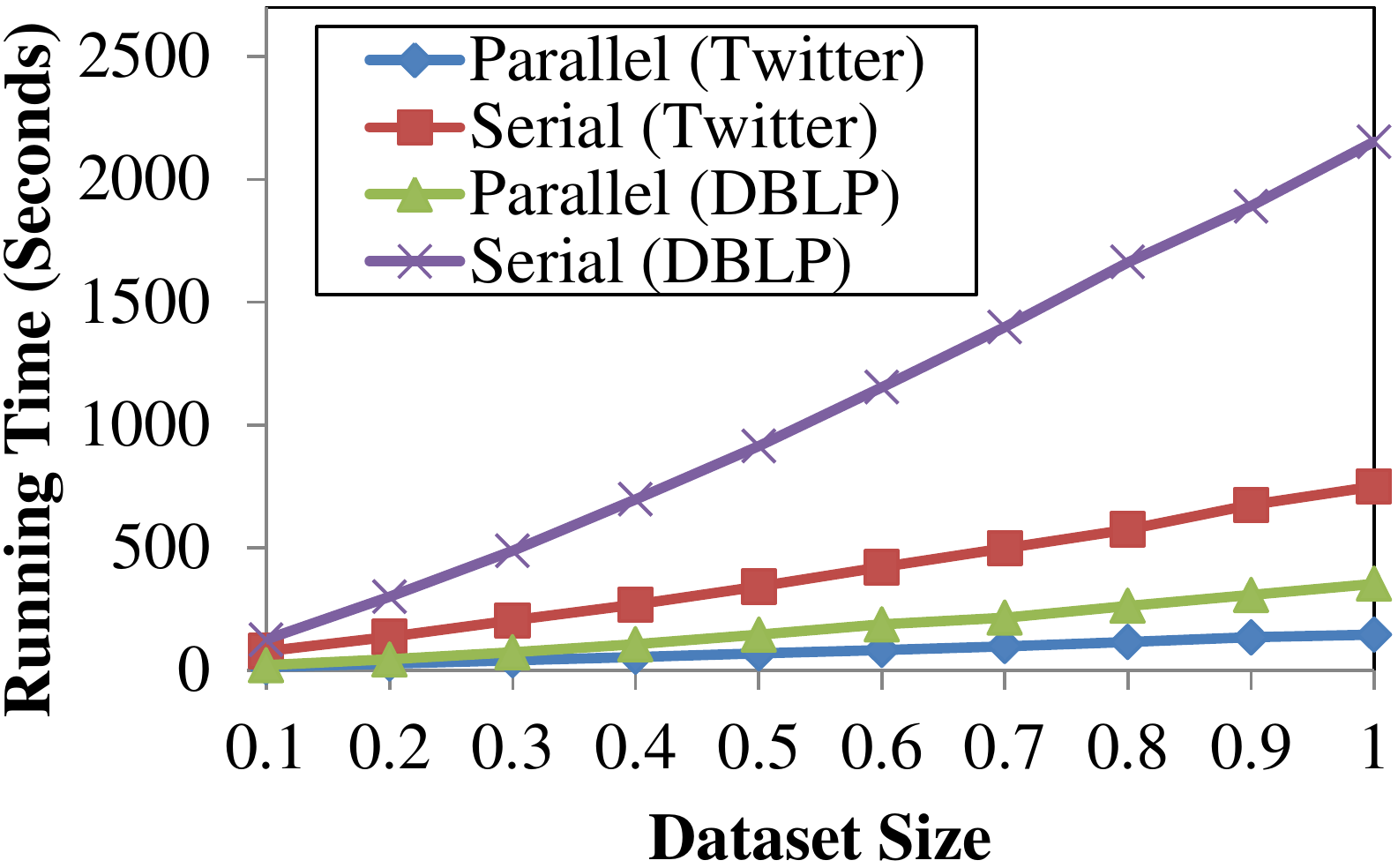}} \hspace{1mm}
    \subfigure[Time vs. \#(CPU Cores)]{\label{fig:cpus}
        \includegraphics[width=0.48\linewidth]{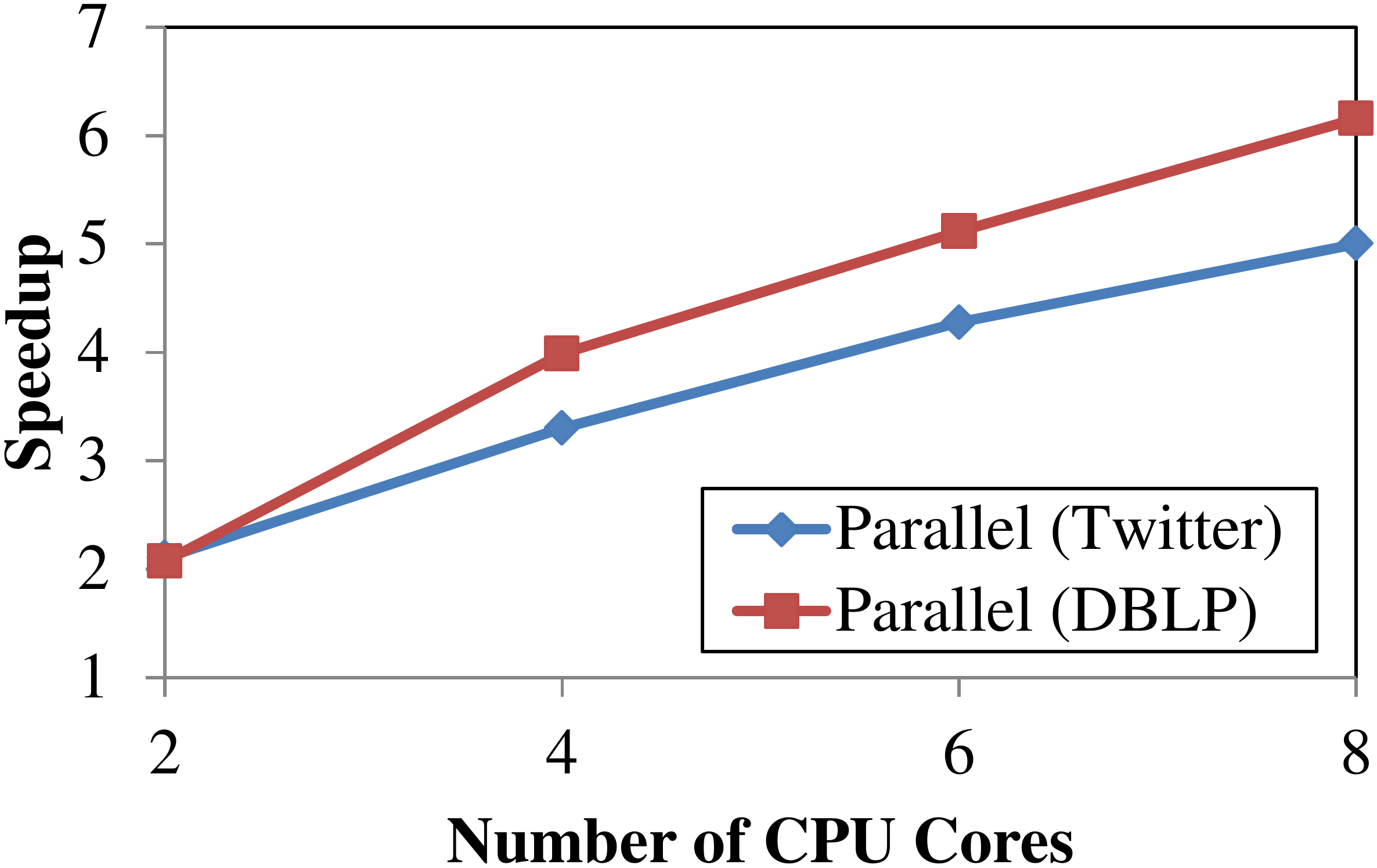}}
\end{tabular}
\vspace{-0.5cm}
\caption{Results of scalability.} \label{fig:mt}
\end{figure}

In Fig.~\ref{fig:scalability}, we first show that our training time (per iteration, Alg.~\ref{alg.inference}'s steps 3--10) scales linearly to the data set size. Each value $p$ (e.g., $p=0.1$) in the x-axis of Fig.~\ref{fig:scalability} indicates that we randomly sample $(p \times 100)$ percents of the total documents, friendship links and diffusion links for experiments. We repeat ten times and report the average training time. We set $|C|=150$ and $|Z|=150$. Different $|C|$ and $|Z|$ can change the absolute training time, but they do not change the linearity of our training time to the data set size. Moreover, we also show that our multithread parallelization achieves up to 4.5$\times$ and 5.7$\times$ speedup over the serial implementation in Twitter and DBLP respectively, by using eight CPU cores. 

\begin{figure}
\small
\begin{tabular}[t]{c} \hspace{-0.6cm}
   \subfigure[Estimated workload]{ \label{fig:estimated_workload}
        \includegraphics[width=0.52\linewidth]{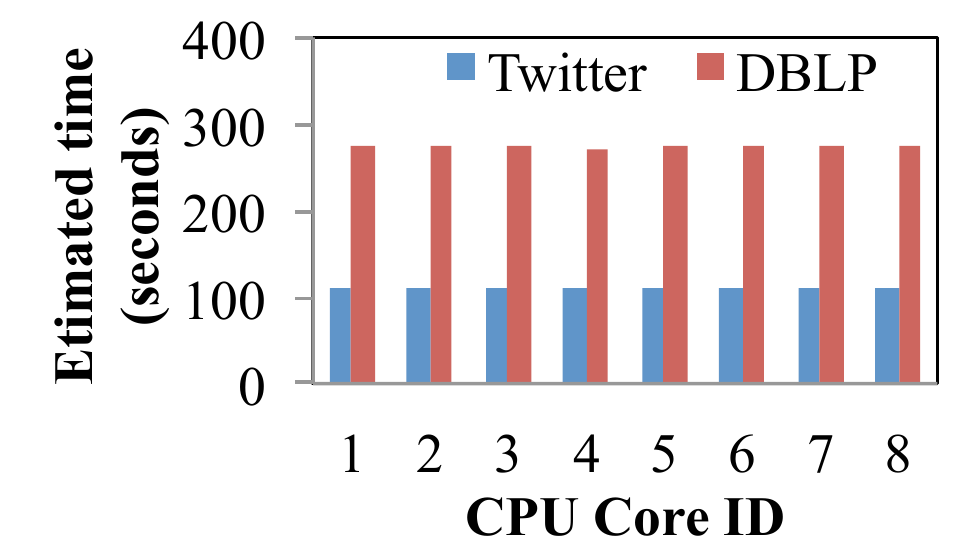}} 
    \subfigure[Actual running time]{\label{fig:actural_runtime} \hspace{-2mm}
        \includegraphics[width=0.52\linewidth]{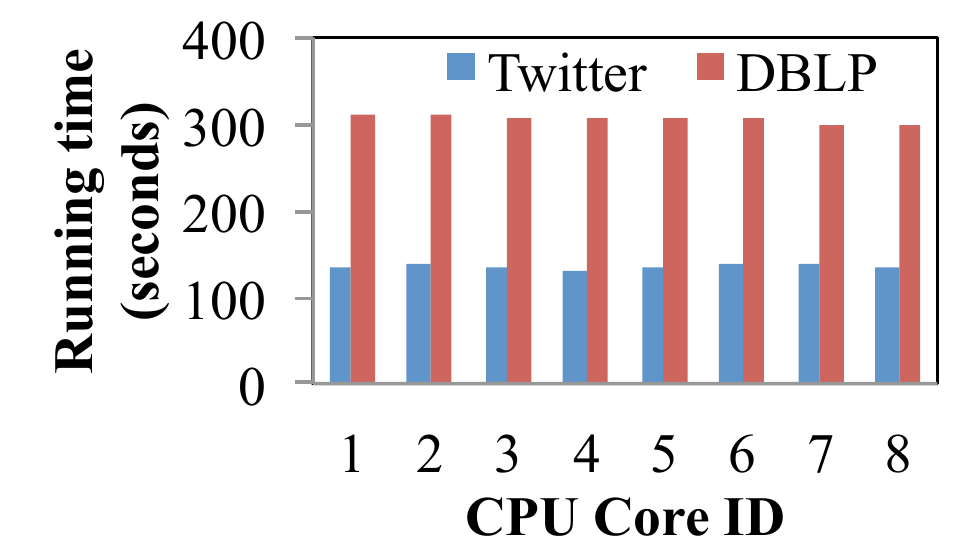}}
\end{tabular}
\vspace{-0.5cm}
\caption{Workload balancing for CPU Cores.} \label{fig:load_balance}
\end{figure}

In Fig.~\ref{fig:cpus}, we plot the speedup with different number of CPU cores in parallelization. 
Generally, the speedup increases as using more CPU cores. We observe that the speedup for DBLP data set is bigger than that of Twitter. 
That maybe because compared with Twitter, DBLP users tend to have less diverse topics in their documents. This makes each \emph{data segment} (defined in Sect.~\ref{subsec:modelscalability}) more likely to have a single topic, which greatly reduces the inter-dependency between the data segments. In Fig.~\ref{fig:load_balance}, we also plot the estimated workload and the actual running time of each CPU core. As we can see, our parallelization design achieves good workload balancing. 

\section{Conclusion}
In this paper, we study a novel problem of community profiling. Community profiling is different from community detection, and its goal is to characterize each community with both its internal profile and external profile. Community profiling also enables many new community-level applications. 
The difficulty of community profiling is largely overlooked. Thus we propose a CPD model, which novelly identifies and addresses three key challenges, including the inter-dependency with community detection, the heterogeneity of social observations and the nonconformity of user behaviors. We also develop a scalable inference algorithm for the CPD model; it scales linearly with the data set size, and we further parallelize it with multithreading. 
In our experiments, we use two public, large-scale, real-world data sets. 
We extensively evaluate CPD in terms of its community detection quality and its community profile quality. 
We verify that our model design well addresses the three challenges. 
We also show that CPD outperforms the state-of-the-art baselines in a number of tasks, including community detection, friendship link prediction, community-aware diffusion, profile-driven community ranking and content profile evaluation. 

In future, we plan to explore other types of user information for defining the profiles, such as user attributes, and user sentiments.

\section{Acknowledgement}
We thank the support of: National Natural Science Foundation of China (No. 61502418), Zhejiang Provincial Natural Science Foundation (No. LQ14F020002), Research Grant for Human-centered Cyber-physical Systems Programme at Advanced Digital Sciences Center from Singapore’s Agency for Science, Technology and Research (A*STAR), and NSF Grant IIS 16-19302. Any opinions, findings, and conclusions or recommendations expressed in this publication are those of the author(s) and do not necessarily reflect the views of the funding agencies.

\balance

\bibliographystyle{abbrv}
\bibliography{sigproc}  


\end{document}